\documentclass{emulateapj}

\usepackage{amsmath}
\usepackage{amssymb}
\usepackage{amsfonts}
\usepackage{amstext}
\usepackage{graphicx}
\usepackage{fancybox}
\usepackage{enumitem}
\usepackage[usenames,dvipsnames]{color}
\usepackage{pstricks}
\usepackage{bm}
\usepackage{soul, color}
\usepackage{latexsym}
\usepackage{pifont}
\usepackage{shorttoc}
\usepackage{subfigure}
\usepackage{textcomp} 
\usepackage{hyperref}
\usepackage{bm}
\usepackage{lipsum}
\usepackage{threeparttable}
\usepackage{epsfig}

\def\aap{A\&A}
\def\mnras{MNRAS}
\def\apj{ApJ}
\def\apjs{ApJS}
\def\apjl{ApJL}
\def\nat{Nature}

\def\aj{AJ}
\def\araa{ARAA}

\newcommand{\no}[1]{}

\newcommand{\hs}{\hspace{1mm}}

\newcommand{\fesch}{$f^{H2}_{\rm esc}$}

\def\lsim{~\rlap{$<$}{\lower 1.0ex\hbox{$\sim$}}}

\def\gsim{~\rlap{$>$}{\lower 1.0ex\hbox{$\sim$}}}

\shorttitle{3-cm Masers from DCBHs}
\shortauthors{Dijkstra et al.}

\begin{document}

\title{3-cm Fine Structure Masers: A Unique Signature of Supermassive Black Hole Formation via Direct Collapse in the Early Universe}

\author{Mark Dijkstra\altaffilmark{1}, Shiv Sethi\altaffilmark{2} \& Abraham Loeb\altaffilmark{3}}
\affil{$^1$Institute of Theoretical Astrophysics, University of Oslo,
P.O. Box 1029 Blindern, N-0315 Oslo, Norway}
\affil{$^2$Raman Research Institute, C V Raman Avenue, Bangalore--560080, India}
\affil{$^3$Harvard Smithsonian Center for Astrophysics, 60 Garden St, Cambridge, MA, 02138, USA}
\altaffiltext{1}{mark.dijkstra@astro.uio.no}
\altaffiltext{2}{sethi@rri.res.in}

\begin{abstract}
The direct collapse black hole (DCBH) scenario describes the isothermal collapse of a pristine gas cloud directly into a massive, $M_{\rm BH}=10^4\hbox{--}10^6 M_{\odot}$ black hole. In this paper we show that large HI column densities of primordial gas at $T\sim 10^4$ K with low molecular abundance---which represent key aspects of the DCBH scenario---provide optimal conditions for pumping of the $2p$-level of atomic hydrogen by trapped Ly$\alpha$ photons. This Ly$\alpha$ pumping mechanism gives rise to inverted level population of the $2s_{1/2}-2p_{3/2}$ transition, and therefore to stimulated fine structure emission at $\lambda=3.04$ cm (rest-frame). We show that simplified models of the DCBH scenario amplify the CMB by up to a factor of $\sim 10^5$, above which the maser saturates. Hyperfine splitting of the 3-cm transition gives rise to a characteristic broad (FWHM$\sim$ tens of MHz in the observers frame) asymmetric line profile. This signal subtends an angular scale of $\sim 1\hbox{--}10$ mas, which translates to a flux of $\sim$ 0.3-3 $\mu$Jy, which is detectable with ultra-deep surveys being planned with SKA1-MID. While challenging, as the signal is visible for a fraction of the collapse time of the cloud, the matching required physical conditions imply that a detection of the redshifted 3-cm emission line would provide direct evidence for the DCBH scenario.
\end{abstract}


 
\section{Introduction}
\label{sec:intro}

The origin of supermassive black holes (SMBHs) in the Universe---especially those at $z\gsim 6$ \citep[e.g.][]{Fan01,Willott09,Mortlock11,Venemans13}---is still not understood \citep[see e.g][]{Volonteri12,Haiman13}. The `Direct Collapse Black Holes' (DCBH) scenario provides an intriguing possibility in which primordial gas inside dark matter halos with $T_{\rm vir} \geq 10^4$~K collapses {\it directly} into a $\sim 10^4\hbox{--}10^6 M_{\odot}$ black hole, without any intermediate star formation. DCBH formation requires primordial gas to collapse isothermally at $T\sim 10^4$ K \citep[e.g.][]{BL03}, which has been shown to prevent fragmentation \citep[e.g.][]{Li03,Omukai05}. 

Isothermal collapse at $T\sim 10^4$~K is possible if primordial gas is prevented from forming molecular hydrogen (H$_2$, which acts as a gas coolant) during collapse. Formation of H$_2$ is prevented if the gas is bathed in a strong photo dissociating background \citep[e.g.][]{BL03}. Photodissociation occurs via ({\it i}) direct photodissociation by Lyman-Werner (LW) radiation (E$=10.2\hbox{--}13.6$~eV), or ({\it ii}) indirect photodissociation through photo-detachment of H$^-$, which catalyses the formation of H$_2$, by infrared (IR) radiation (E$> 0.76$ eV). DCBH formation therefore requires the collapse of primordial gas inside atomically cooling halos, exposed to an intense Lyman-Werner and/or IR radiation fields. It is also 
possible to achieve near isothermal collapse with $T \sim 10^4$ K if the primordial haloes
are threaded with (comoving) primordial magnetic fields of nano-Gauss strength \citep[][]{Sethi10,Borm13}. 

The DCBH formation process is a remarkably complex problem, which involves the hydrodynamics of gas collapsing from $\sim$ a few kpc size down to event horizon of the black hole \citep[see][for hydrodynamical simulations of this process]{BL03,Latif13,Fernandez14,Regan14,LV15}. In addition, the `critical' intensity of radiation background, indicated with $J_{\rm crit}$, that is needed to keep the gas free of $H_2$ depends on the precise spectral shape of the radiation background \citep[including even the X-ray band, see e.g.][]{Omukai05,Shang10,WG12,Sugimura14,Inayoshi15,Agarwal15}. Because $J_{\rm crit}$ typically greatly exceeds that of the cosmic background, DCBH formation requires a nearby galaxy to boost the intensity of the local radiation field \citep{D08,Agarwal12,Visbal14,Visbalpair}. This implies that the LW-radiation field is not isotropic, which also affects the value of $J_{\rm crit}$ \citep{Regan14}. Furthermore, the spectral dependence of $J_{\rm crit}$ implies that $J_{\rm crit}$ depends on the stellar populations of the nearby galaxy \citep{Agarwal15}. Finally, the close proximity to a star forming galaxies makes it more complicated to prevent enrichment of the gas by feedback-driven outflows originating from the nearby galaxy. It has been shown that small uncertainties in these processes can lead to orders of magnitude changes in the predicted number density of DCBHs \citep{LW,D14}.

Because of the large theoretical uncertainties associated with DCBH formation, it is extremely valuable to have observational signposts on this process. \citet{Agarwal13} have presented predictions for the broad-band colours of DCBH host galaxies, under the assumption that the spectrum emitted by the accretion disk surrounding the DCBH is a multi-colored disk. Under this assumption, DCBH host galaxies are characterised by blue UV slopes ($\beta \sim -2.3$), which are similar to those predicted for metal poor, young stars. \citet{Kohei} more recently estimated the Ly$\alpha$ luminosity from the accretion flow onto the central black hole to be comparable to that of known Ly$\alpha$ emitting galaxies.

The goal of this paper is to focus on the two strongest {\it fine structure} lines of atomic hydrogen which include\footnote{We adopted the notation $nL_{J}$, where $n$ is the principle quantum number, $L$ denotes the electron's {\it orbital} angular momentum, and $J$ denotes the {\it total} (orbital + spin) quantum number.}: ({\it i}) the $2p_{1/2}\rightarrow 2s_{1/2}$ transition at $\lambda \approx 27$ cm ($\nu_{\rm ul}=1.1$ GHz), and ({\it ii}) the $2s_{1/2}\rightarrow 2p_{3/2}$ transition at $\lambda \approx 3.04$cm ($\nu_{\rm ul}=9.9$ GHz) \citep[see e.g.][for more details]{Wild52,Ershov87,Dennison05,Sethi07,D08}. It has long been realised that scattering of Ly$\alpha$ photons can ``pump'' the $2p$-level, and give rise to stimulated 3-cm emission \citep[e.g.][]{Pottasch,FP61,Ershov87}. These early studies focussed on pumping of the $2p-$level in nearby HII regions, where Ly$\alpha$ scattering is limited by dust, and not nearly effective enough to give rise to stimulated 3-cm emission \citep[][]{MB72}. However, the DCBH scenario is associated with primordial (i.e. dust-free) gas cloud with extremely large column densities of atomic hydrogen ($N_{\rm HI}\gsim 10^{22}-10^{24}$ cm$^{-2}$, also see Paccuci \& Ferrara 2015) at $T\sim 10^4$ K. These conditions are ideal for Ly$\alpha$ photons to be both produced, and undergo a large number of scattering events, both of which are favourable for pumping the $2p$-level of atomic hydrogen. Moreover, modelling of the relevant radiative processes is simplified in the absence of star formation \& stellar feedback in the DCBH formation scenario. It is therefore highly timely to study the fine structure signatures of gas clouds directly collapsing into a black hole.

The 3-cm fine structure masers studied in this paper thus arise due to extremely efficient pumping by Ly$\alpha$ absorption of the excited upper level of the 3-cm transition. This pumping mechanism distinguishes the 3-cm maser from the more recently studied radio recombination line (RRL) masers \citep{S96,Spaans97}. RRL masers can arise in the $n\alpha$ transitions (in which $\Delta n=1$) for $n\gg 1$ due to collisional pumping: the efficiency with which electrons can excite these transitions increases as $\propto n^{4-5}$, while the spontaneous decay rate between these transitions decreases as $n\propto n^{-5}$ \citep[see][for a more extended discussion]{S96,Spaans97}. 

The outline of this paper is as follows: \S~\ref{sec:model} describes our simplified model of the direct-collapse black hole scenario, the relevant processes that determine the $2s_{1/2}$ and $2p_{3/2}$ level populations and their fine-structure signatures. We present our main results in \S~\ref{sec:signal}. We discuss the detectability of stimulated fine structure emission in \S~\ref{sec:detection}. We discuss our model assumptions in \S~\ref{sec:discuss} before finally presenting our main conclusions in \S~\ref{sec:conc}. For completeness, in the concordance cosmology ($\Omega_{\rm m}=0.3,\Omega_{\Lambda}=0.7,h=0.7$) the mass of a dark matter halo with virial temperature of $T_{\rm vir}=10^4$ K is  $M_{\rm tot}=10^8(\mu/0.6)^{-3/2}([1+z]/11)^{-3/2}$ $M_{\odot}$, which has a virial radius of $r_{\rm vir}=1.8([1+z]/11)^{-1}(M_{\rm tot}/10^8 M_{\odot})^{1/3}$ kpc \citep{BL01}. The average number density of hydrogen atoms/nuclei at virialization is $\bar{n}= 0.048([1+z]/11)^3$ cm$^{-3}$. 

\begin{figure*}
\begin{center}
\epsfig{file=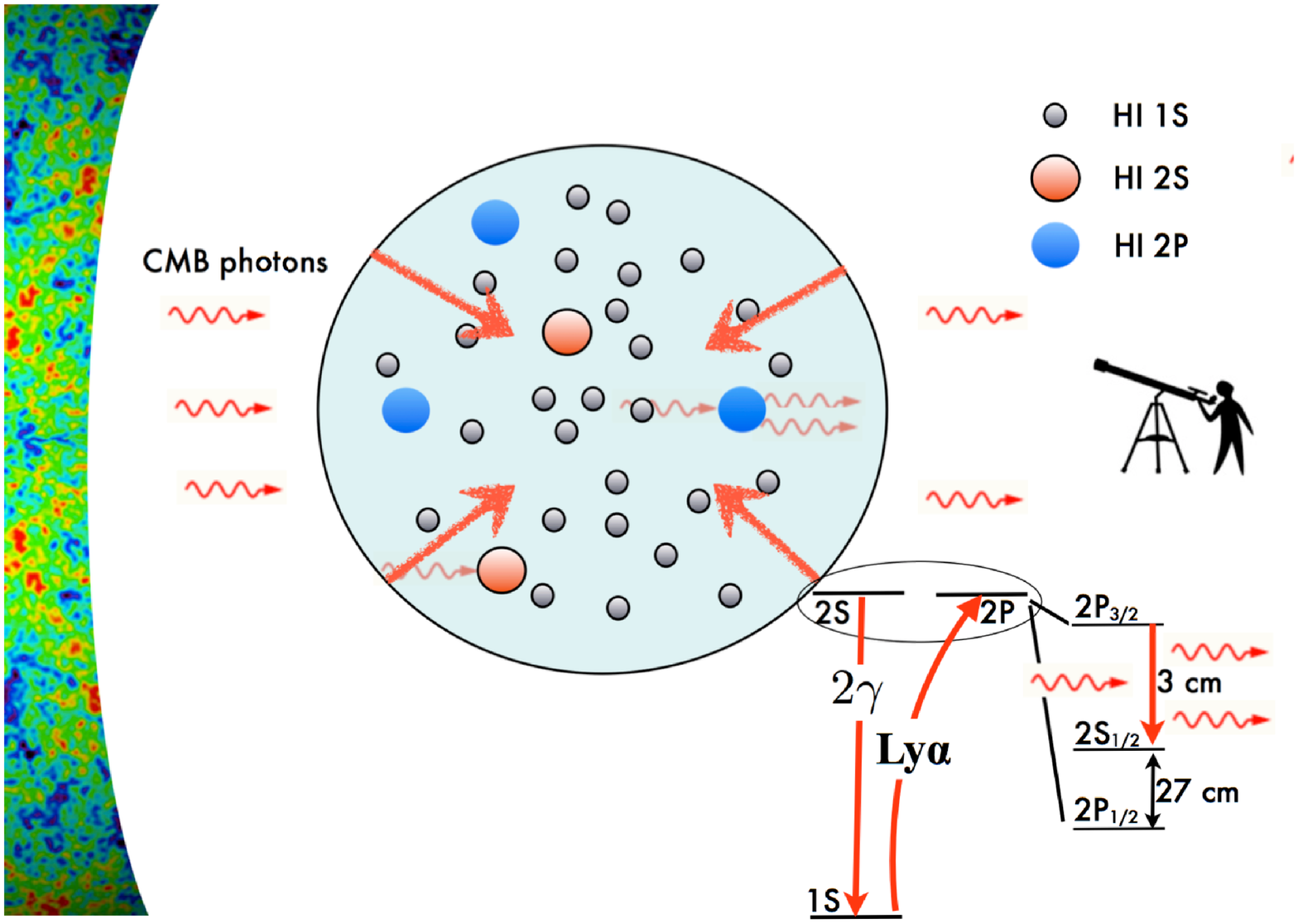,angle=270,width=12.5truecm}
\caption{Schematic representation of the calculations in this paper. A gas cloud collapses directly into a `direct collapse black holeÕ (DCBH). In the DCBH scenario, the gas cloud consists solely of atomic hydrogen gas (indicated by {\it circles}). The column density of atomic hydrogen, $N_{\rm HI}$, is typically huge and can exceed $N_{\rm HI}=10^{24}$ cm$^{-2}$. Depending on the temperature of the cloud, a small fraction of atomic hydrogen is in the first excited state ($n=2$, indicated by the {\it larger, colored circles}), which has two fine structure transitions at $1.1$ GHz ($\lambda=27$ cm, $2p_{1/2}\rightarrow 2s_{1/2}$) and $9.9$ GHz ($\lambda=3$ cm, $2s_{1/2}\rightarrow 2p_{3/2}$).These transitions are indicated in the lower right corner of the figure. In the case where the $2p_{3/2}$ is overpopulated with respect to $2s_{1/2}$, CMB radiation passing through the cloud induces stimulated emission in this transition. The {\it thick red arrows} indicate the maser cycle: Ly$\alpha$ pumps the $2p_{3/2}$ level. The CMB induces stimulated emission $2p_{3/2}\rightarrow 2s_{1/2}$, which is followed by decay to the ground state via two-photon emission.} 
\label{fig:scheme}
\end{center}
\end{figure*}

\section{Model}
\label{sec:model}

\subsection{Geometry }
Our analysis is limited to spherically symmetric gas clouds with a uniform density. These simplifying assumptions offer us a clear view on the relevant radiative processes that determine whether the masing conditions exist. In particular, they allow us to treat the radiative transfer of Ly$\alpha$ photons analytically, which represents a major computational advantage. Under these assumptions the cloud is fully characterised by a single number density, $n$.  For a given gas mass $M_{\rm gas}$, this gives a cloud radius $R$. We discuss in \S~\ref{sec:discuss} how our main results are expected to be affected by these simplifying assumptions. 

Hydrodynamical simulations indicate that that the gas density profile is closer to isothermal (i.e $\rho(r) \propto r^{-2}$, e.g. Shang et al. 2010, Pacucci \& Ferrara 2014). The gas density increases towards the centre of the collapsing gas cloud, which may lead to the formation of a quasi-star, a supermassive star, or a direct-collapse black hole in the centre of the cloud.
 We therefore also consider models in which the gas cloud contains a central source of ionising radiation, which represents a scenario in which a central black hole has already formed. Accretion rates onto the central black hole can be up to $\sim 0.1 M_{\odot}$ yr$^{-1}$ (e.g. Latif \& Volonteri 2015), which can power a central source with a luminosity exceeding $10^{44}$ erg s$^{-1}$.
 We do not consider the hydrodynamic impact of the central source on the gas. Instead, we focus on the emission properties of gas with properties that were favourable for the DCBH scenario, i.e. the gas is pristine, no fragmentation occurred, and no stars have formed. 

\subsection{The Fine Structure Signal}

We study the impact of the collapsing halo on CMB photons passing through them (see Fig~\ref{fig:scheme}). In general, a gas cloud with uniform density changes the CMB intensity, $I_{\nu}$, by an amount
\begin{equation}
I_{\nu}(s)=I_{\nu,0}{\rm e}^{-\kappa_{\nu} s}+\frac{j_{\nu}}{\kappa_{\nu}}\left(1- {\rm e}^{-\kappa_{\nu} s}\right),
\label{eq:RTsolution}
\end{equation} at a distance $s$ into the cloud. Here, $I_{\nu,0}$ denotes the intensity before entering the cloud, $\kappa_{\nu}$ denotes the opacity of the cloud (in cm$^{-1}$), and $j_{\nu}$ denotes its volume emissivity (in erg s$^{-1}$ cm$^{-3}$ Hz$^{-1}$). Eq~(\ref{eq:RTsolution}) implicitly assumes that the cloud is static. To account for gas motion, we would need to Doppler boost into the frame of the gas at $s$. However, we will show in \S~\ref{sec:discuss} that the fine structure cross-sections are extremely broad in frequency, and that this Doppler boost can be safely ignored. For the relevant results in this paper, the second term on the right-hand-side of Eq~(\ref{eq:RTsolution}) is much smaller than $I_{\nu}$ (see Appendix~\ref{app}) in which case Eq~\ref{eq:RTsolution} simplifies to 
\begin{equation}
I_{\nu}(s)=I_{\nu,0}{\rm e}^{-\kappa_{\nu} s}.
\label{eq:RTsolution2}
\end{equation}

The opacity through any transition is given by \citep[see e.g.][Eq~1.78]{RL79}
\begin{eqnarray}
\kappa_{\nu}=\kappa(\nu)&=&\frac{h\nu_{\rm ul}B_{\rm ul}}{4\pi  \Delta \nu_{\rm ul}}\times \Big{(}\frac{g_u}{g_l}n_{\rm l}-n_{\rm u} \Big{)}\phi(\nu), \\ \nonumber
j_{\nu}=j(\nu)&=&\frac{h\nu_{\rm ul}}{4\pi\Delta \nu_{\rm ul}}n_uA_{\rm ul}\phi(\nu),
\label{eq:kj}
\end{eqnarray} where $B_{\rm ul}=\frac{c^2}{2 h\nu^3}A_{\rm ul}$,  $A_{\rm ul}$ denotes the Einstein A-coefficient for the transition from the higher energy state `u' to the lower energy state `l'. The energy difference between the two states is given by $\Delta E=h\nu_{\rm ul}$. The factors $g_{\rm l}$ and $n_{\rm l}$ ($g_{\rm u}$ and $n_{\rm u}$) denote the statistical weight and number density of atoms in the lower (upper) energy state. The function $\phi(\nu)$ denotes the line profile function (also known as the Voigt function), which is normalized to unity through $\frac{1}{\Delta \nu_{\rm ul}}\int d\nu \phi(\nu)=1$, in which $\Delta \nu_{\rm ul} \equiv \nu_{\rm ul}\sqrt{\frac{2kT}{m_{\rm p}c^2}}$. 

We consider the $2p_{1/2}\rightarrow 2s_{1/2}$ (for which $g_{\rm l}=2$, $g_{\rm u}=2$, $A_{\rm ul}=1.60 \times 10^{-9}$ s$^{-1}$) and the $2s_{1/2}\rightarrow 2p_{3/2}$ transition (for which $g_{\rm l}=2$, $g_{\rm u}=4$, $A_{\rm ul}=8.78 \times 10^{-7}$ s$^{-1}$). We assume that the atoms in the $2p$ state are divided between $2p_{1/2}$ and $2p_{3/2}$ following their statistical weight. That is, for the 9.9~GHz transition we have $n_{\rm l}=n_{\rm 2s}$ and $n_{\rm u}=n_{\rm 2p}\times 2/3$, while for the 1.1~GHz we have $n_{\rm l}=n_{\rm 2p}\times 1/3$ and $n_{\rm u}=n_{\rm 2s}$. We justify adopting this assumption in \S~\ref{sec:discuss}.

We write the line center total optical depth $\tau^{\rm FS}_0$ through the center of the collapsing gas cloud of radius $R_{\rm cl}$---in {\it both} fine structure transitions---as 
\begin{equation}
\tau^{\rm FS}_0\equiv 2R_{\rm cl}\kappa_0=R_{\rm cl}\frac{\lambda^2A_{\rm ul}}{\pi}\times \Big{(}\frac{g_u}{g_l}n_{\rm l}-n_{\rm u} \Big{)} \frac{1}{A_{\alpha}},
\label{eq:tau}
\end{equation} where $A_{\alpha}=6.25 \times 10^8$ s$^{-1}$ is the Einstein coefficient for the $2p\rightarrow 1s$ transition. This expression does not contain the temperature-dependent width $\Delta \nu_{\rm D}$, because the Voigt parameter\footnote{Both fine structure transitions involve transitions between $2s$ and $2p$ states, both of which have finite lifetimes. Under these conditions, the relevant Voigt parameter equals $a_{v}=\frac{A_{\alpha}+A_{\rm 2s1s}}{4\pi \Delta \nu_{\rm D}}$, where  $\Delta \nu_{\rm D}=\frac{\nu_{ul}}{c}\sqrt{\frac{2kT}{m_p}}$ (see Eq 10.74, Rybicki \& Lightman 1979). Since $A_{\alpha} \gg A_{\rm 2s1s}$, we have $a_{v}=\frac{A_{\alpha}}{4\pi \Delta \nu_{\rm D}}$ to high accuracy.} that is present in the Voigt function, $\phi(\nu)$ is $a_{\rm v}=\frac{A_{\alpha}}{4 \pi \Delta \nu_{\rm D}}\approx 117(T_{\rm gas}/10^4)^{-1/2}$ for the $3$cm-transition, and $a_{\rm v}\approx 13(T_{\rm gas}/10^4)^{-1/2}$ for the $27$cm-transition \citep[][]{D08}. The line profile function evaluated at line center is  \citep[see e.g.][]{CS09}
\begin{equation}
\phi(0)=\frac{1}{\sqrt{\pi}}\exp(a^2_{\rm v}){\rm erfc} (a_{\rm v})\approx \frac{1}{\pi a_{\rm v}} =\frac{4 \Delta \nu_{\rm ul}}{A_{\alpha}},
\end{equation} where we have used that ${\rm erfc}(a_{\rm v}) \rightarrow \frac{e^{-a_{\rm v}^2}}{a_{\rm v}\sqrt{\pi}}$ when $a_{\rm v} \gg 1$. The factor $\Delta \nu_{\rm ul}$ that enters $\phi(0)$ cancels the one that is present in the expression for $j(\nu)$. The opacity through the fine structure transitions is therefore independent of temperature, which is because the fine structure transitions
 have large intrinsic  spectral width (see Fig~\ref{fig:signal}) and thermal broadening has no impact.

\subsection{The Level Populations of HI}
\label{sec:level}

\subsubsection{The $2s$--Level}
\label{sec:2s}

We first list the processes that populate the $2s$-level, and quantify the rates at which these occur. We then list the processes that {\it de-}populate the $2s$ level. The processes that populate the $2s$ level include

\begin{itemize}[leftmargin=*]
\item {\bf Collisional excitation from the ground state.} The total rate at which collisional excitation of a hydrogen atom in the ground state with a free electron leaves the hydrogen atomic its $2s$ state is $n_en_{{\rm 1s}}C_{1s2s}$ (in cm$^{-3}$ s$^{-1}$), where $n_e$ denotes the number density of free electrons, and where
\begin{equation}
C_{\rm lu}=8.63\times 10^{-6}T^{-1/2}\langle \Omega_{lu}\rangle \exp \Bigl( -\frac{\Delta E_{lu}}{k_BT} \Bigr) \hs {\rm cm}^{3}\hs{\rm s}^{-1}.
\end{equation} 
Here, $\langle \Omega_{lu} \rangle$ denotes the `velocity averaged collision strength' of the $1s\hbox{--}2s$ transition. We adopt $\langle\Omega_{lu}\rangle=0.27$, which corresponds to the value appropriate at $T=10^4$ K \citep[with a very weak temperature dependence, see][]{Scholz90}. 
\item {\bf Indirect photo excitation by higher Lyman-series photons.} The rate at which the $2s$ level is populated as a result of absorption of a Lyman series photon by HI in its ground state, which subsequently radiatively cascades down to the $2s$ state is $\Gamma_{\rm Lyn}n_{\rm 1s}P_{np2s}$ (in cm$^{-3}$ s$^{-1}$). We can safely ignore this term in the gas clouds we are considering. This is because higher order Lyman series photons scatter only a small number of times before being converted into lower energies photons, with suppresses their scattering rate relative to that of Ly$\alpha$ by orders of magnitude (see Dijkstra et al. 2008b for an extended discussion).

\item {\bf Recombination into the $2s$-state}. The rate at which the $2s$ level is populated as a result of recombination of an electron and proton into the 2s state (either through direct recombination into the $2s$ state, or via some intermediate higher energy state followed by a radiative cascade into $2s$, is $\alpha_{2s}n_en_p$ (in cm$^{-3}$ s$^{-1}$). For case-B recombination, $f_{\rm 2s}\sim 32\%$ of all recombination events will result in an atom populating the $2s$ term \citep{Spitzer51,Dennison05}. That is, $\alpha_{\rm 2s}=f_{\rm 2s}\alpha_{\rm rec,B}$, where $\alpha_{\rm B}$ denotes the case-B recombination coefficient, which we take from \citet{HG97}. 
\item {\bf Collisional transitions $2p \rightarrow 2s$.} The $2s$ level can be populated as a result of a collisions between hydrogen atoms in their $2p$ with a free proton (collisions with electrons are $\sim 10$ times less efficient). The rate for this process is $n_pn_{2p}C_{2p2s}$ (in cm$^{-3}$ s$^{-1}$), with the rate coefficient $C_{\rm 2p2s}=1.8 \times 10^{-4}$ cm$^{3}$ s$^{-1}$ \citep[e.g.][and references therein]{Dennison05}.
\item {\bf Direct Radiative transitions $2p \rightarrow 2s$.} Finally, the $2s$ can be populated via direct radiative transitions, either via spontaneous or via CMB-induced transitions. This rate is $n_{\rm 2p}(A_{2p2s}+\Gamma^{\rm CMB}_{2p2s})$. Spontaneous radiative transitions are only allowed for $2p_{3/2} \rightarrow 2s_{1/2}$, for which $A_{2p2s}=8.78 \times 10^{-8}$ s$^{-1}$. Under our assumption that 2/3 of all atoms in $2p$-state are  in the $2p_{3/2}$ level (which we justify in \S~\ref{sec:discuss}), this rate becomes $\frac{2}{3}n_{\rm 2p}A_{2p2s}$. The CMB induces transitions $2p \rightarrow 2s$ in two ways: ({\it i}) stimulated emission from the $2p_{3/2}$ state, and ({\it ii}) absorption from the $2p_{1/2}$ state. The rate at which this happens is  $\frac{2}{3}n_{\rm 2p}\Gamma^{\rm CMB}_{2p_{3/2}2s}+\frac{1}{3}n_{\rm 2p}\Gamma^{\rm CMB}_{2p_{1/2}2s}$. This can be recast\footnote{Eq~(\ref{eq:kj}) shows that the cross-section for absorption from some lower state `$l$' to an upper state `$u$' equals the cross-section for stimulated emission from the upper to the lower level, but multiplied by the ratio of statistical weights $\frac{g_u}{g_l}$. The stimulated emission rate (per atom in the $2p_{3/2}$ state) from $2p_{3/2} \rightarrow 2s_{1/2}$ is therefore half that of the absorption rate (per atom in the $2s_{1/2}$ state) from $2s_{1/2} \rightarrow 2p_{3/2}$.} as $\frac{2}{3}n_{\rm 2p}\frac{1}{2}\Gamma^{\rm CMB}_{2s2p_{3/2}}+\frac{1}{3}n_{\rm 2p}\Gamma^{\rm CMB}_{2s2p_{1/2}}=\frac{1}{3}n_{\rm 2p}9.3\times 10^{-6}(1+z)$, where we adopted the rates from Eq~A3 in Hirata (2006). In short, we define $\Gamma^{\rm CMB}_{2p2s}\equiv 3.1\times 10^{-6}(1+z)$ s$^{-1}$. It is also clear the CMB induced transitions dominate over spontaneous transitions.
\end{itemize}

Processes that {\it de}populate the $2s$-level include
\begin{itemize}[leftmargin=*]
\item {\bf Collisional transitions $2s \rightarrow 2p$.} The rate at which HI atoms leave the $2s$ as a result of collisionally induced transitions to the $2p$ state is $C_{2s2p}n_{p}n_{\rm 2s}$ (in cm$^{3}$ s$^{-1}$), where $C_{\rm 2s2p}=3\times C_{\rm 2p2s}=5.3 \times 10^{-4}$ cm$^{3}$ s$^{-1}$.
\item {\bf Two-photon Decay $2s \rightarrow 1s$.} The rate at which hydrogen atoms leave the $2s$ state as a result of a direct radiative transition to the ground state (by emitting two photons) is $n_{2s}A_{2s1s}$ (in cm$^{3}$ s$^{-1}$). The Einstein $A$-coefficient for such a transition is $A_{\rm 2s1s}=8.25$ s$^{-1}$. 
\item {\bf Direct Radiative transitions $2s\rightarrow 2p$}. Finally, the $2s$ can be depopulated via the direct radiative transitions mentioned above. This rate is $n_{\rm 2s}(A_{2s2p}+\Gamma^{\rm CMB}_{2s2p})$. Spontaneous radiative transitions are only allowed for $2s_{1/2} \rightarrow 2p_{1/2}$, for which $A_{\rm 2s2p}=1.60 \times 10^{-9}$ s$^{-1}$. The CMB again induces transitions $2s \rightarrow 2p$ in two ways: ({\it i}) stimulated emission from the $2s_{1/2}$ state, and ({\it ii}) absorption into the $2p_{3/2}$ state. Following the arguments given above, we can write the rate at which this happens as  $n_{\rm 2s}\Gamma^{\rm CMB}_{2s_{1/2}2p_{1/2}}+n_{\rm 2s}\Gamma^{\rm CMB}_{2s_{1/2}2p_{3/2}}\equiv \Gamma^{\rm CMB}_{2s2p}n_{\rm 2s}$. We note that $\Gamma^{\rm CMB}_{2s2p}=3\Gamma^{\rm CMB}_{2p2s}$.
\end{itemize}

The equilibrium solution is given by

\begin{equation}
A+Bn_{\rm 2p}=Cn_{\rm 2s},
\label{eq:rate2s2}
\end{equation} with 
\begin{eqnarray}
A=&n_{\rm 1s}[n_eC_{1s2s}+\Gamma_{\rm Lyn}P_{np2s}]+\alpha_{2s}n_en_p, \\ \nonumber
B=&n_pC_{2p2s}+A_{2p2s}+\Gamma^{\rm CMB}_{2p2s}, \\ \nonumber
C=&A_{2s1s}+A_{2s2p}+C_{2s2p}n_{p}+\Gamma^{\rm CMB}_{2s2p}.
\label{eq:ABC}
\end{eqnarray}

\subsubsection{The $2p$--Level}

The equilibrium solution of the $2p$ state can be written in a simplified form  very similar to Eq~\ref{eq:rate2s2}:
\begin{equation}
D+En_{\rm 2s}=Fn_{\rm 2p},
\label{eq:rate2p2}
\end{equation} where

\begin{eqnarray}
D =&n_{\rm 1s}\Gamma_{\alpha}\\ \nonumber
E =& n_pC_{2s2p}+\Gamma^{\rm CMB}_{2s2p}\\ \nonumber
F =& A_{\alpha}+C_{2p2s}n_{p}+A_{2p2s}+\Gamma^{\rm CMB}_{2p2s}.
\label{eq:DEF}
\end{eqnarray} The structure of these terms is similar to those describing the population and depopulation of the $2s$ levels. The most significant difference is in the $D$-term, which only contains the term $\Gamma_{\alpha}n_{\rm 1s}$. This term denotes the rate at which Ly$\alpha$ photons are absorbed, i.e. scattered, by hydrogen atoms in the ground state into the $2p$ state. Because of the large optical depth of the gas cloud to Ly$\alpha$ photons, Ly$\alpha$ photons typically scatter many times off different atoms before escaping from the cloud. The Ly$\alpha$ scattering rate is therefore boosted compared to the Ly$\alpha$ production rate as
\begin{eqnarray}
\Gamma_{\alpha} = ({\rm Ly}\alpha\hs{\rm production \hs rate\hs {\rm per }\hs H\hs  atom})\times \nonumber \\ \times ({\rm number \hs of \hs scattering \hs events \hs {\rm per } \hs Ly}\alpha\hs{\rm photon})= \nonumber \\=\big[n_{e}C_{\rm 1s2p}+\Gamma_{\rm Lyn}P_{np2p}+\alpha_{2p}n_en_p/n_{\rm 1s}\big]\langle N_{\rm scat}\rangle,
\end{eqnarray} where $\alpha_{\rm 2p}=f_{\rm 2p}\alpha_{\rm rec, B}$ with $f_{\rm 2p}=1-f_{\rm 2s}\approx 0.68$ \citep[see][for a derivation]{Dijkreview}, and where we adopt $\langle \Omega_{1s2p} \rangle=0.47$ in the expression for $C_{\rm 1s2p}$ \citep{Scholz90}. Finally, $\langle N_{\rm scat}\rangle$ denotes the mean number of times a Ly$\alpha$ photon scatters before it escapes from the cloud. 

The Ly$\alpha$ scattering process can be described by diffusion in both real and frequency space \citep[e.g.][]{Adams72,Harrington73,Neufeld90}. The diffusion in frequency space, combined with the strong frequency dependence of the Ly$\alpha$ absorption cross-section, causes $\langle N_{\rm scat} \rangle$ to scale with the line center optical depth of the medium to Ly$\alpha$ photons as $\langle N_{\rm scat} \rangle \propto \tau_{{\rm Ly}\alpha,0}$ (Adams 1972, as opposed to the $\langle N_{\rm scat} \rangle \propto \tau^2$ dependence that is expected for a random walk in real space only). For Ly$\alpha$ photons emitted in the center of a uniform, static sphere we have $\langle N_{\rm scat} \rangle\approx k_1 \tau_{0}$ with $k_1=0.6$ \citep{D06}. In our case the gas cloud is collapsing and is therefore not static. However, the infall velocity is expected to be close to the circular velocity of the dark matter halo hosting the gas cloud, which is $v_{\rm circ} \sim 10$ km s$^{-1}$. This infall velocity is comparable to the thermal velocity of the gas, and gas motions do not modify our estimate at all \citep[e.g.][and see \S~\ref{sec:discuss} for a more quantitative discussion]{SS06}.

The line center optical depth $\tau_{0} \propto N_{\rm HI}=2n_{\rm HI}R_{\rm cl}\propto n^{2/3}M^{1/3} \propto n^{2/3}$ at fixed $M$, and therefore increases as the cloud continues its collapse. Importantly, $\langle N_{\rm scat} \rangle$ does not increase indefinitely with $\tau_{0}$ for four main reasons (in Appendix~\ref{app:other} we discuss a few more reasons that are less important):
\begin{enumerate}[leftmargin=*] 
\item At increasingly high densities,  collisional de-excitation from the $2p$ state becomes more probable, which would result in the destruction of the Ly$\alpha$ photon. The probability that this occurs at any scattering event is given by $p_{\rm dest}=\frac{n_pC_{\rm 2p2s}}{n_pC_{\rm 2p2s}+A_{\alpha}}$. At a given number density $n$, we therefore expect Ly$\alpha$ photons not to scatter more than $\approx p_{\rm dest}^{-1}$ times. 
\item Gas clouds collapsing into a DCBH do contain small amounts of molecular hydrogen, with $f_{H_2}\equiv n_{H_2}/n_{\rm H}\sim 3\hbox{--}5\times 10^{-9}$ \citep[e.g.][]{Shang10,Latif15}. Molecular hydrogen has two transitions that lie close to the Ly$\alpha$ resonance: ({\it a}) the $v=1-2P(5)$ transition, which lies $\Delta v=99$ km s$^{-1}$ redward of the Ly$\alpha$ resonance, and ({\it b}) the $1-2R(6)$ transition which lies $\Delta v=15$ km s$^{-1}$ redward of the Ly$\alpha$ resonance. Vibrationally excited $H_2$ may therefore convert Ly$\alpha$ photons into photons in the $H_2$ Lyman bands  \citep[][and references therein]{Neufeld90}, and thus effectively destroy Ly$\alpha$. \citet{Neufeld90} provides an expression for the escape fraction of Ly$\alpha$ photons from a static slab whose line centre optical depth from slab centre to slab edge is $\tau_0$, which we denote with \fesch$(\tau_0)$. We reproduce the full expression for \fesch$(\tau_0)$ \hs in Appendix~\ref{app:h2}. We take the simple approach and assume  that $\langle N_{\rm scat} \rangle \rightarrow \langle N_{\rm scat} \rangle\times$\fesch$(\tau_0)$, i.e. only those photons that escape contribute to the scattering rate. The contribution from photons that are destroyed by molecular hydrogen is ignored. 

\item When a non-negligible fraction of hydrogen atoms is in the first exited state, Ly$\alpha$ photons can photoionize these excited atoms. The photoionisation cross-section from the $n=2$ by Ly$\alpha$ photons is $\sigma^{{\rm Ly}\alpha}_{\rm ion}=5.8 \times 10^{-19}$ cm$^{2}$ (e.g. Cox 2000, p 108). Because Ly$\alpha$ photons scatter so frequently, their total path through the cloud is increased by a factor of $B=(a_v\tau_0/\sqrt{\pi})^{1/3}\approx 12(N_{\rm HI}/10^{20}\hs{\rm cm^{-2}})^{1/3}(T/10^4\hs{\rm K})^{-1/3}$ \citep{Adams75}. The total optical depth for photoionisation from the $n=2$ state that Ly$\alpha$ photons experience equals $\tau_{\rm ion}^{{\rm Ly}\alpha}=B[n_{2p}+n_{2s}]\sigma_{\rm ion}R_{\rm cl}\equiv B \tau_{\rm ion}$. We can substitute $\tau_0=\langle N_{\rm scat} \rangle/k_1$ into the expression for $B$, and get a maximum number of scattering events that a Ly$\alpha$ photon can undergo by setting $\big(\frac{a_v N^{\gamma,{\rm max}}_{\rm scat}}{\sqrt{\pi}k_1}\big)^{1/3}\equiv 1$. This translates to $N_{{\rm scat}}^{\gamma,{\rm max}}=\frac{\sqrt{\pi}k_1}{a_v\tau^3_{\rm ion}}$. Because $\tau_{\rm ion}$ depends on the number density of hydrogen atoms in the $2s$ and $2p$ states (which we are trying to solve for), this complicates the analysis slightly. We first ignore this process. After computing $n_{2p}$ and $n_{2s}$, we will verify whether this assumption was justified.

\item Finally, Ly$\alpha$ photons are destroyed in the maser cycle itself (see Fig~\ref{fig:scheme}). This becomes important when the maser saturates. We do not include this effect in our calculations, because it would require us to simultaneously solve for the level populations, and the amplified CMB through the cloud. Instead, we verify whether ignoring this effect was justified in \S~\ref{sec:3cmsignal}.
\end{enumerate}
We therefore compute $\langle N_{\rm scat} \rangle$ as

\begin{equation}
\label{eq:nscat}
\langle N_{\rm scat} \rangle=\min\Big{(}k_1\tau_{{\rm Ly}\alpha,0}f_{\rm esc}^{H2}, p^{-1}_{\rm dest}\Big{)}.
\end{equation} 

\subsubsection{Solving the Rate Equations}

For a primordial gas cloud with uniform density, the number density of hydrogen nuclei (i.e. free protons plus neutral hydrogen atoms), $n$, equals

\begin{equation}
n=\frac{3(1-Y_{\rm He})M_{\rm gas}}{4\pi R_{\rm cl}^3 m_{\rm p}}\approx 100 \big(\frac{M_{\rm gas}}{10^{7}\hs M_{\odot}} \big)\big(\frac{R_{\rm cl}}{0.1\hs{\rm kpc}}\big)^{-3}\hs{\rm cm}^{-3}
\end{equation} where $Y_{\rm He}=0.24$ denotes the primordial Helium mass fraction.

To determine the fraction of hydrogen atoms in the first excited state, we need the gas temperature, and we therefore need to specify heating mechanisms. We assume that the gas is heated by two different processes:
\begin{itemize}[leftmargin=*]

\item {\bf Gravitational Heating.} This corresponds to the heating associated with the contraction of the cloud, which converts gravitational binding energy into kinetic energy, which in turn is converted into heat (e.g. Haiman et al. 2000). The total gravitational heating rate is thus
\begin{eqnarray}
& H^{\rm grav}  =  \frac{dU_{\rm bind}}{dt}=\frac{3GM_{\rm gas}^2}{5R^2}\dot{R}  \\ \nonumber
& \approx 1.7\times 10^{38}\hs{\rm erg/s} \left( \frac{M_{\rm gas}}{10^7\hs M_{\odot}}\right)^2  \left( \frac{R}{100\hs{\rm pc}}\right)^{-2}  \left( \frac{\dot{R}}{10\hs{\rm km/s}}\right), \\ \nonumber
&\approx 1.7\times 10^{38}\hs{\rm erg/s} \left( \frac{M_{\rm gas}}{10^7\hs M_{\odot}}\right)^2  \left( \frac{n}{100\hs{\rm cm}^{-3}}\right)^{2/3} \left( \frac{\dot{R}}{10\hs{\rm km/s}}\right) ,
\label{eq:gravheat}
\end{eqnarray} where we used that $U_{\rm bind}=-\frac{3GM^2_{\rm gas}}{5R}$. We therefore assumed that the  dark matter does not contribute to the gravitational potential, which is appropriate when the gas has collapsed to the high densities that we consider in this paper. 

\item {\bf Radiative Heating.} In case the gas is collapsing onto an inplace DCBH (see \S~\ref{sec:model}), the radiation from the accretion disk can photoionize the gas cloud. When the total (maximum) recombination rate of the cloud, $\dot{N}^{\rm max}_{\rm rec}=\alpha_{\rm B}n^2\frac{4}{3}\pi R^3_{\rm cl}$, exceeds the rate at which the accretion disk produces ionising photons, $\dot{N}_{\rm ion}$, then the accretion disk will not be able to fully ionise the gas cloud. In this case, the cloud consists of an HII region of radius $R_{\rm ion}$, which is surrounded by neutral gas (which extends out to $R_{\rm cl}$). X-ray photons emitted by the accretion disk can penetrate this neutral gas and heat it. For the production rate of ionising photons we assume the DCBH is accreting at Eddington luminosity and that its spectrum is identical to that of unobscured, radio-quiet quasars\footnote{More specifically, this assumes a broken power-law spectrum of the form $f_{\nu} \propto \nu^{-0.5}$ for $1050\hs$\AA$<\lambda<1450\hs$\AA, and $f_{\nu} \propto \nu^{-1.5}$ for $\lambda < 1050$ \AA \hs \citep{Bolton11}.}. Under these assumptions, $\dot{N}_{\rm ion}=6.5\times 10^{53} \left(\frac{M_{\rm BH}}{10^6\hs M_{\odot}} \right)\hs{\rm s}^{-1}$. 

The total maximum recombination rate of a cloud is $\dot{N}^{\rm max}_{\rm rec}=\frac{4}{3}\pi R^3_{\rm cl}n^2\alpha_{\rm B}(T)=\frac{M_{\rm gas}n\alpha_{\rm B}(T)}{\mu m_{\rm p}}$. We obtain the `critical' density, $n_{\rm crit}$, beyond which the gas cloud cannot be kept ionised by setting $\dot{N}^{\rm max}_{\rm rec}=\dot{N}_{\rm ion}$. This critical density equals

\begin{eqnarray}
n_{\rm crit} & =& \frac{\mu m_{\rm p} \dot{N}_{\rm ion}}{M_{\rm gas} \alpha_{\rm B}(T)} \\ \nonumber
& \approx & 240\left(\frac{M_{\rm BH}}{10^6\hs M_{\odot}} \right)\left( \frac{M_{\rm gas}}{10^7\hs M_{\odot}}\right)^{-1}\left(\frac{T}{10^4\hs {\rm K}} \right)^{0.7}\hs {\rm cm}^{-3},
\label{eq:ncrit}
\end{eqnarray} where the $T-$dependence is due to the recombination coefficient. For $n > n_{\rm crit}$ the radiative heating is important. The total column density of HI through the cloud is 
\begin{eqnarray}
N_{\rm HI}=n^{2/3}\left(\frac{3M_{\rm gas}}{4\pi \mu m_{\rm p}}\right)^{1/3}\left[1 -\left(\frac{n_{\rm crit}}{n} \right)^{1/3}\right]= \\ \nonumber
=nR_{\rm cl}\left[1 -\left(\frac{n_{\rm crit}}{n} \right)^{1/3}\right].
\label{eq:columnH}
\end{eqnarray} We use this to compute $\tau_{{\rm Ly}\alpha,0}$ and hence $\langle N_{\rm scat} \rangle$.

When the radiative heating is important, we assume that a fraction $f_{\rm heat}$ of the X-ray luminosity goes into heating. The total heating rate is then
\begin{eqnarray}
H^{\gamma}& = & f_{\rm heat}f_{\rm X}L_{\rm edd} \\ \nonumber
&=&1.3\times 10^{42}\left( \frac{f_{\rm heat}}{0.1}\right)\left(\frac{f_{\rm X}}{0.1}\right)\left( \frac{M_{\rm BH}}{10^6 \hs M_{\odot}}\right)\hs {\rm erg}\hs{\rm s}^{-1}, 
\label{eq:gheat}
\end{eqnarray} where the choice $f_{\rm X}\sim 10\%$ is based on observations of luminous quasars for which 10\% of  their total bolometric luminosity is in the 0.5-10 keV band \citep[e.g.][]{Marconi04,Lusso12}. The choice $f_{\rm heat}=10\%$ is arbitrary, but reflects the possibility that not all X-ray photons  are absorbed in the gas, and that not all their energy goes into heating of the gas. 

\end{itemize} 

For a given heating process and density, we compute the equilibrium temperature $T_{\rm eq}$ for which radiative cooling balances this heating, i.e.
\begin{equation}
H=C_{\rm tot}(T_{\rm eq},x_{\rm e}),
\label{eq:teq}
\end{equation} where $C_{\rm tot}(T_{\rm eq},x_{\rm e})$ denotes the total cooling rate. For the gas-temperatures encountered in this paper ($T=[0.6-1.2]\times 10^4$ K, see Appendix~\ref{app:intermediate} for details) cooling is completely dominated by collisional excitation of atomic hydrogen (see e.g. Fig~ 1 of Thoul \& Weinberg 1995), but we have included other cooling processes as well. Eq~(\ref{eq:teq}) simultaneously gives us the ionised fraction $x_{\rm e}\equiv \frac{n_e}{n}=\frac{n_p}{n}$ and $T_{\rm eq}$.\\

At this point, for a given $n$ and heating mechanism, we have $n_e=n_p$ and $T$. We can then solve for $n_{\rm 2s}$ and $n_{\rm 2p}$ by combining Eq~(\ref{eq:rate2s2}) and Eq~(\ref{eq:rate2p2}) to give
\begin{eqnarray}
\label{eq:n2p}
\hspace{20mm}n_{\rm 2p}=\frac{CD+EA}{CF-EB}, \\ \nonumber
\hspace{20mm}n_{\rm 2s}=\frac{A+Bn_{\rm 2p}}{C}.
\end{eqnarray} where $A,...,F$ were defined in Eq~(\ref{eq:DEF}) and Eq~(\ref{eq:ABC}).

After we have computed $n_{\rm 2s}$ and $n_{\rm 2p}$, we verify whether $N_{{\rm scat}}^{\gamma,{\rm max}}>\langle N_{\rm scat} \rangle$, where $N_{{\rm scat}}^{\gamma,{\rm max}}$ is the maximum number of scattering events a Ly$\alpha$ photon undergoes before photoionizing a hydrogen atom from its $n=2$ state. If $N_{{\rm scat}}^{\gamma,{\rm max}}>\langle N_{\rm scat} \rangle$ then our calculation is accurate. If on the other hand, $N_{{\rm scat}}^{\gamma,{\rm max}}<\langle N_{\rm scat} \rangle$, then we allowed Ly$\alpha$ photons to scatter too frequently. As we show below, the number densities $n_{\rm 2s}$ and $n_{\rm 2p}$ both depend linearly on $\langle N_{\rm scat} \rangle$. If we force $\langle N_{\rm scat} \rangle$ - and therefore $n_{\rm 2s}$ and $n_{\rm 2p}$- to be suppressed by a factor of $x$ ($x<1$), then $N_{{\rm scat}}^{\gamma,{\rm max}} \propto \tau_{\rm ion}^{-3} \propto (n_{\rm 2p}+n_{\rm 2s})^{-3} \propto x^{-3}$. In this case, we want to increase $N_{{\rm scat}}^{\gamma,{\rm max}}$ such that it equals $\langle N_{\rm scat} \rangle$. We achieve this by choosing $x=(N_{{\rm scat}}^{\gamma,{\rm max}}/\langle N_{\rm scat} \rangle)^{1/3}$, and rescale $n_{\rm 2p} \rightarrow x n_{\rm 2p}$. We then recompute $n_{\rm 2s}$ by applying Eq~\ref{eq:n2p}. We found that this correction is only necessary over a restricted range of densities, and that even it only corrects our predicted optical depths at the tens of per cent level.

\section{Results}
\label{sec:signal}

\subsection{The Number Densities $n_{\rm 2s}$ and $n_{\rm 2p}$}
\label{sec:num2s2p}

\begin{figure}
\begin{center}
\epsfig{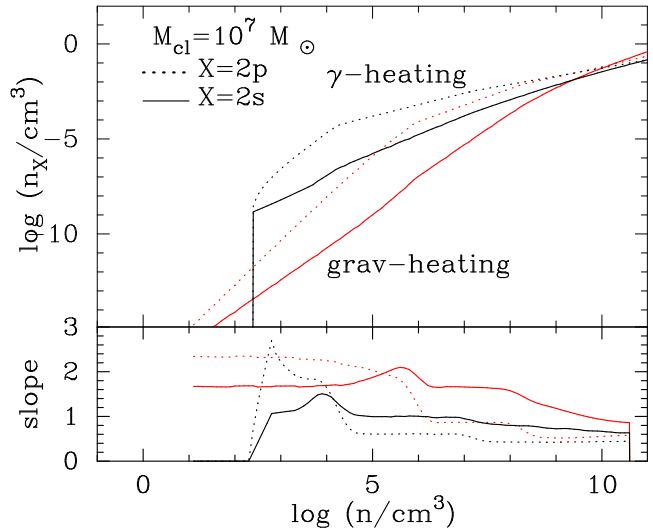}
\vspace{0mm}
\caption[]{The {\it top panel} shows the dependence of $n_{\rm 2s}$ ($n_{\rm 2p}$) on $n$ as {\it solid lines} ({\it dotted lines}), for the cases of gravitational heating ({\it red lines}) and radiative heating ({\it black lines}). The {\it lower panel} shows the slope of these lines. Both $n_{\rm 2p}$ and $n_{\rm 2s}$ are $\ll n_{\rm 1s}$, i.e. atoms in the first excited state account for only a tiny fraction of hydrogen. The $2p$-state is overpopulated with respect to the $2s$-state at all densities, for both heating mechanisms in spite of the much shorter life-time of atoms in the $2p$ state. The reason the $2p-$level is overpopulated is that Ly$\alpha$ scattering boosts the excitation rate into the $2p$-state enormously. This overpopulation of the $2p-$level leads to stimulated emission of the $2p_{3/2} \rightarrow 2s_{1/2}$ transition at $\lambda=3$ cm. The number densities go to zero at $n \lsim 240$ cm$^{-3}$ in the radiative heating case as we expect the gas cloud to be fully ionised below these densities (see text).} 
\label{fig:n}
\end{center}
\end{figure}

Figure~\ref{fig:n} shows both $n_{\rm 2s}$ ({\it solid lines}) and $n_{\rm 2p}$ ({\it dotted lines}) as a function of $n$ for the case of gravitational heating ({\it red lines}), and radiative heating ({\it black lines}). The {\it bottom-panel} shows the density dependence of the slope d$\log n_{\rm 2s/2p}/$d$\log n$ of each of the lines shown in the {\it top panel}. The density-dependence of $n_{\rm 2s}$ and $n_{\rm 2p}$ is easy to understand. We discuss both separately:

\begin{itemize}[leftmargin=*]

\item {\bf 2s-Level.} The {\it lower red solid line} that represents the gravitational heating scenario shows that $n_{\rm 2s} \propto n^{1.67}$ up to $\log (n/{\rm cm}^{-3})\sim 4.5$. This is because at these densities the $2s$ state is populated primarily via collisional excitation from the ground state, and depopulated via two-photon emission, and thus $n_{\rm 2s} \propto n_{\rm e}n_{\rm 1s}/A_{\rm 2s1s}$. Gas cooling is dominated by collisional excitation of atomic hydrogen, and the total cooling rate therefore scales as $L_{\rm cool} \propto V n_e n_{\rm 1s}$, where $V$ denotes the total volume of the cloud. The cooling rate balances the total gravitational heating rate, $H^{\rm grav} \propto n^{2/3}$. We therefore can see that $n_e n_{\rm 1s} \propto L_{\rm cool}/V = H^{\rm grav}/V \propto n^{5/3}$, which explains the slope of the $n_{\rm 2s}\hbox{--}n$ relation in Figure~\ref{fig:n}.

For intermediate density ($\log (n/{\rm cm}^{-3}) \sim 4.5\hbox{--}6.0$) Figure~\ref{fig:n} shows a steepening of the relation, which is because here the $2s$ level is populated via collisionally induced transitions $2p \rightarrow 2s$. The coupling to the $2p$ level boosts the density dependence of $n_{\rm 2s}$, because of the stronger density-dependence of $n_{\rm 2p}$ at these densities (as shown by the {\it dotted lines}, we explain this density-dependence below).

At higher densities ($\log (n/{\rm cm}^{-3}) \gsim 6.0$) Figure~\ref{fig:n} shows that the slope of the $n_{\rm 2s}\hbox{--}n$ relation is almost the same as in the low-density regime. The $2s$-level is still populated primarily via collisional induced $2p \rightarrow 2s$ transitions, which locks the $n_{\rm 2s}$ evolution to $n_{\rm 2p}$. However, at $\log n \gsim 6.0$ collisional destruction of Ly$\alpha$ photons becomes important which changes the density-dependence of $n_{\rm 2p}$. Finally, at very high densities the $n_{\rm 2s}$ is limited by collisionally induced transitions of the form $2s \rightarrow 2p$.

The {\it upper solid line} represents the radiative heating scenario. The density-dependence of this curve can be understood in a similar way. The  two main differences are that ({\it i}) for $n < n_{\rm crit} \sim 240$ cm$^{-3}$ (see Eq~\ref{eq:ncrit}) the cloud is fully ionised and $n_{\rm 2s}\sim 0$, and ({\it ii}) the total radiative heating rate $H^{\gamma}$ does not depend on density, which causes $n_e n_{\rm 1s} \propto L_{\rm cool}/V = H^{\gamma}/V \propto n$. This explains why the slope of the $n_{\rm 2s}\hbox{--}n$ line is flatter. The spike in the slope near $n\sim 240$ cm$^{-3}$ is numerical. Finally, the enhanced number density of protons makes collisional deexcitation more important at lower densities in the radiative heating models, which is why the slope approaches $\sim 0.8\hbox{--}0.9$ at lower densities.

\item {\bf 2p-Level.} Figure~\ref{fig:n} also shows that the $n$-dependence of $n_{\rm 2p}$ is different. We first discuss the gravitational heating scenario (the lines that represent the radiative heating models can be understood similarly). The $2p$ level is primarily populated via absorption of Ly$\alpha$ photons. The production of Ly$\alpha$ photons is dominated by collisional excitation. We showed in the discussion above that the collisional excitation rate increases as $n_e n_{\rm 1s} \propto n^{5/3}$. For low densities (now $\log n/({\rm cm}^{-3}) < 4.0$) the  total average number of scattering events per Ly$\alpha$ photon increases as $\langle N_{\rm scat} \rangle \propto \tau_0 \propto N_{\rm HI} \propto n^{2/3}$. The total Ly$\alpha$ scattering rate therefore increases as $\propto n^{7/3}$, which is the density-dependence of $n_{\rm 2p}$ shown in Figure~\ref{fig:n}.

At densities $\log n/({\rm cm}^{-3}) \sim 4.0\hbox{--}6.0$ the slope of the $n\hbox{--}n_{\rm 2p}$  flattens from $n_{\rm 2p}\propto n^{7/3}$ to $n_{\rm 2p}\propto n^{1.2}$ (with most change in the range $\log n\sim 5.0\hbox{--}6.0$). This change in the slope arises because $\langle N_{\rm scat} \rangle$ becomes limited by $H_2$, and \fesch \hs drops below unity for $\log n/({\rm cm}^{-3}) > 4.0$. At higher densities ($\log n/({\rm cm}^{-3}) \gsim 6.0$), collisional deexcitation becomes important, and we have $\langle N_{\rm scat} \rangle \propto n_{\rm p}^{-1}$. The proton number density $n_{\rm p}$ only increases $\propto n$ at fixed temperature. Because the cloud temperature (slightly) decreases with $n$, $n_{\rm p} \propto n^{0.8}$ (see Fig~\ref{fig:xp} in Appendix~\ref{app:intermediate}), and we have $n_{\rm 2p} \propto n^{5/3}\langle N_{\rm scat} \rangle \propto n^{5/3}n_{\rm p}^{-1}\propto n^{5/3-0.8} \propto n^{0.8-0.9}$, which is what is shown in Figure~\ref{fig:n} up to $\log n/({\rm cm}^{-3}) \sim 8.0$. Finally, at the highest densities $\log n/({\rm cm}^{-3}) \gsim 8.0$, $n_{\rm 2p} \propto n^{0.5-0.6}$, which is because the number of times Ly$\alpha$ photons scatter ($\langle N_{\rm scat} \rangle$) is limited by the optical depth to photoionisation from the $n=2$ state (see Fig~\ref{fig:nscat} in Appendix~\ref{app:intermediate}). In this case $n_{\rm 2p} \propto N_{{\rm scat}}^{\gamma,{\rm max}} n^{5/3} \propto  \tau_{\rm ion}^{-3} n^{5/3} \propto n_{\rm 2p}^{-3}n^{5/3}$, which therefore implies that $n^4_{\rm 2p} \propto n^{5/3}$, i.e. $n_{\rm 2p} \propto n^{5/12} \approx n^{0.4}$, which is close to what we see.
\end{itemize}

Initially, the $2p$ state is significantly overpopulated relative to the $2s$ state. The simplest explanation for this is that the $2p$ population rate in boosted by $\langle N_{\rm scat} \rangle$ compared to $2s$-population rate. This boost $\langle N_{\rm scat} \rangle$ lies in the range $\log \langle N_{\rm scat} \rangle \sim 8-11$ (see Fig~\ref{fig:nscat} in Appendix~\ref{app:intermediate}), and overcompensates for the enormously shorter natural life-times of the $2p$ state ($t=A_{\alpha}^{-1}\sim 10^{-9}$ s) compared to the $2s$ state ($t=A_{2s1s}^{-1}\sim 0.1$ s). The weaker $n$-dependence of $2p$ at high densities, allows the $2s$ population to `catch-upÕ. At the highest densities we find that $n_{\rm 2p}/n_{\rm 2s} \sim 0.6$ for gravitational heating, and $n_{\rm 2p}/n_{\rm 2s} \sim 2.6$ for radiative heating. Note that with $n\rightarrow \infty$ we expect that $n_{\rm 2p}/n_{\rm 2s} \rightarrow C_{2s2p}/C_{2p2s}=3$. This limit is not reached yet because the $2p$-level is still predominantly populated via Ly$\alpha$ absorption, which elevates $n_{\rm 2p}$ above the value expected for thermal equilibrium.
\begin{figure}
\vbox{\centerline{\epsfig{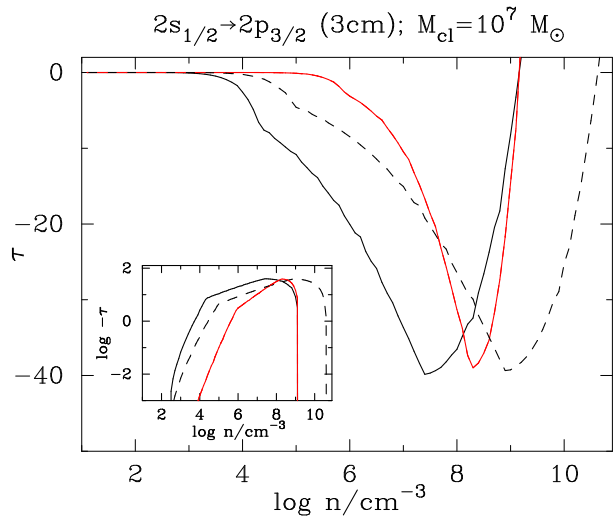}}}
\caption[]{The optical depth $\tau_0$ through the centre of the cloud (total mass $10^7 M_{\odot}$) in the 3-cm ($2s_{1/2} \rightarrow 2p_{3/2}$) fine-structure transition for the case of gravitational heating ({\it red solid line}), and radiative heating with $H^{\gamma}=10^{43}$ erg s$^{-1}$ ({\it black solid line}) and $H^{\gamma}=10^{42}$ erg s$^{-1}$ ({\it black dashed line}). The enhanced population in the $2p-$state (see Fig~\ref{fig:n}) gives rise to considerable negative optical depth. The gas cloud thus acts as a maser amplifying the background CMB. The total amplification depends little on the precise heating mechanism, though this mechanism does determine at what density the maximum amplification is reached (see text for a quantitative analysis of all these plots).} 
\label{fig:tau3cm}
\end{figure}

\begin{figure}
\vbox{\centerline{\epsfig{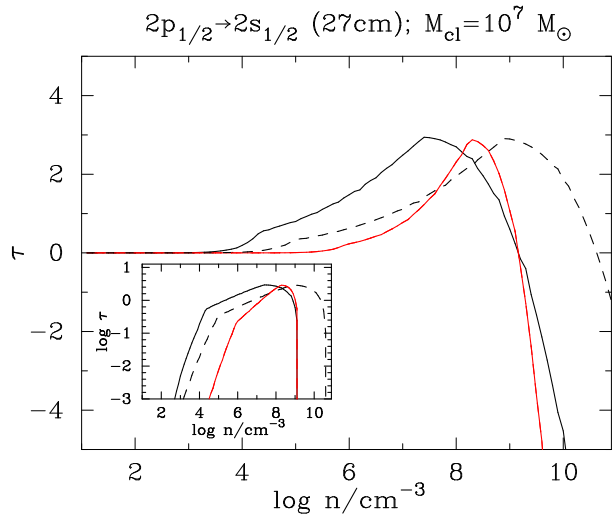}}}
\caption[]{Same as Fig~\ref{fig:tau3cm}, but for the 27-cm ($2p_{1/2} \rightarrow 2s_{1/2}$) fine-structure transition. The precise density dependence of the optical depth closely follows that of the optical depth through the 3-cm transition, but with the sign reversed and its amplitude suppressed. At very high densities $\log n/({\rm cm}^{-3}) \gsim 9$ we get strong stimulated emission where $\tau_0 \propto -n_{\rm 2s}R_{\rm cl} \propto -n^{0.5-0.6}$.} 
\label{fig:tau27cm}
\end{figure}

\subsection{The Line Center Optical Depth $\tau^{\rm FS}_0$}
\label{sec:tau}
Using the values of  $n_{\rm 2s}$ and $n_{\rm 2p}$ at a given $n$ and $T$, we use Eq~\ref{eq:tau} to compute $\tau_0$ in both fine structure transitions. Figure~\ref{fig:tau3cm} shows $\tau_{0}$ for the 3-cm ($2s_{1/2}\rightarrow 2p_{3/2}$ transition), and Figure~\ref{fig:tau27cm} shows $\tau_{0}$ for the 27-cm ($2p_{1/2}\rightarrow 2s_{1/2}$) transition. The {\it solid red lines} show the case of gravitational heating, while the {\it black solid} ({\it black dashed lines}) show cases for radiative heating with $H^{\gamma}=10^{43}$ erg s$^{-1}$ ($H^{\gamma}=10^{42}$ erg s$^{-1}$). These figures show that
\begin{itemize}[leftmargin=*]
\item the optical depth through the 3-cm transition is negative, which is expected given the overpopulation of the $2p$-level as compared to the $2s$ level. 
\item the optical depth through the 3-cm transition is significant irrespective of the heating mechanism: for each model the optical depth reaches\footnote{A quick check of this result can be obtained from results reported in the literature. Field \& Partridge (1961) find that $\tau_{0}({\rm H}\alpha)\approx -700 \tau_{0}(3{\rm cm})$ (for gas at $T=5000$ K). The line centre cross-section for H$\alpha$ is $\sigma_{{\rm H}\alpha,0}\sim 5\times 10^{-13}(T/10^4\hs{\rm K})^{-1/2}$, and therefore $\sigma_{3{\rm cm},0}=7\times 10^{-16}$ cm$^{2}$. Figure~\ref{fig:tau3cm} shows that the minimum for $\tau_{3{\rm cm}}$ is reached for $\log n/({\rm cm}^{-3}) \sim 9.0$ for gravitational heating, for which $n_{\rm 2p} \sim 10^{-2}$ cm$^{-3}$ and $R_{\rm cl} \sim 0.5$ pc. We therefore have $\tau_{3{\rm cm}} \sim 2 R_{\rm cl}n_{\rm 2p}\sigma_{3{\rm cm},0}\sim -30$. This estimate is a factor of $\sim 1.5$ lower than our full calculations (though clearly in the same ball park). Appendix~\ref{app:numbers} discusses this in more detail.} $\tau_0 \sim -40$. The precise heating mechanism only affects the density at which the optical depth reaches its minimum. For low $n$, we have a significant over-population of atoms in the $2p-$level, and $|\tau_0| \propto R_{\rm cl}n_{\rm 2p} \propto n^{-1/3+7/3} \propto n^2$. The gravitational heating model indeed shows this behaviour in the low-density regime (with $\log n/({\rm cm}^{-3}) < 6.0$). At higher densities,. Eq~\ref{eq:tau} shows that the optical depth changes sign when $n_{\rm 2p}=3n_{\rm 2s}$ which occurs at $\log n/({\rm cm}^{-3}) \sim 9$. The radiative heating model with $H^{\gamma}=10^{43}$ erg s$^{-1}$ ({\it black solid line}) behaves very similar, but shifted towards lower densities. The model with an order of magnitude less radiative heating ({\it black dashed line}) is shifted back to higher densities again.
\item the optical depth through the 27-cm transition is positive, and significantly smaller than that in the (absolute value of the) 3-cm transition. The main reason for this is that $A_{ul}$ is smaller by a factor $\sim 550$. The precise density dependence closely follows that of the optical depth through the 3-cm transition, and becomes negative when $n_{\rm 2p}=3n_{\rm 2s}$. At very high densities $\log n/({\rm cm}^{-3}) \gsim 9$ we get strong stimulated emission where $|\tau_0| \propto n_{\rm 2s}R_{\rm cl} \propto n^{0.5-0.6} $.
\end{itemize}

\subsection{The $3.04$-cm Signal}
\label{sec:3cmsignal}

\begin{figure} 
\vbox{\centerline{\epsfig{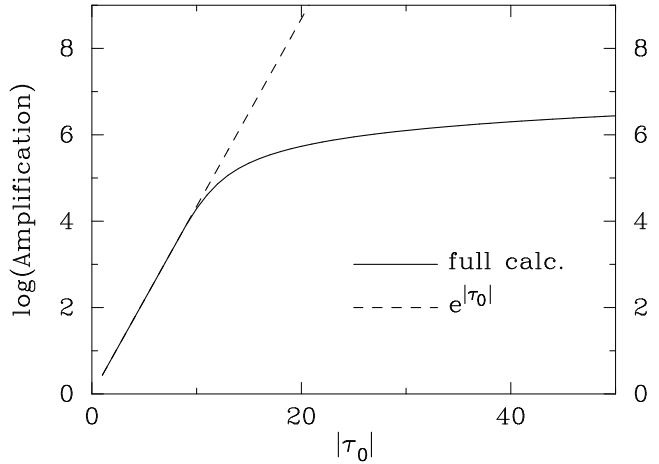}}}
\caption[]{This Figure shows that maser amplification factor as a function of $\tau_0$. The {\it dashed line} shows amplification by a factor of e$^{|\tau_0|}$, while the {\it solid line} shows the actual amplification factor. For $\tau_0 \gsim 10$, the amplified CMB field affects the $2s$ and $2p$ level population, which causes the maser to saturate at amplification factors above $\sim 10^5$.}
\label{fig:amp}
\end{figure}

Eq~\ref{eq:RTsolution2} shows that $I_{\nu}(s)=I_{\nu,0}{\rm e}^{-\tau_0^{\rm FS}}$. The incoming radiation field is that of the CMB. We first recast this expression in terms of the brightness temperature, and compute the brightness temperature difference with the CMB. For $\tau^{\rm FS}_0 < 0$, the brightness temperature is enhanced exponentially:
\begin{equation}
\Delta T_{\rm b}(\nu)=T_{\rm CMB}\left(\exp\left[|\tau^{\rm FS}(\nu)|\right] - 1\right), 
\end{equation} where we have introduced the frequency dependence by replacing $\tau^{\rm FS}_0 \rightarrow \tau^{\rm FS}(\nu)$ (details follow below). The exponental enhancement of the brightness temperature cannot take on arbitrarily large values as the amplified CMB inside the cloud affects the $2p$ and $2s$ level populations, which saturates the maser. The radiative transitions induces by the CMB occur at a rate $\Gamma^{\rm CMB}_{2p2s}\equiv 3.1\times 10^{-6}(1+z)=3.4\times 10^{-5}$ s$^{-1}$. We have repeated our analysis and amplified the CMB intensity by a factor of $B_{\rm CMB}$ and computed $\tau^{\rm FS}_0$ as a function of $B_{\rm CMB}$ (at fixed $n$). Results of this calculation (shown in Appendix~\ref{app:saturation}) show that the maser starts to``shut off'' when $B_{\rm CMB} \gsim 10^5$. At this value of $B_{\rm CMB}$ the $2s$-level is populated at a rate comparable to $A_{\rm 2s1s}$, and at larger values the $2s$ and $2p$ starts to approach the LTE value. The {\it solid line} in Figure~\ref{fig:amp} shows $\Delta T_{\rm b}$ as a function of $\tau_0$, properly taking this effect into account. The {\it dashed line} shows $T_{\rm CMB}\exp\left(|\tau^{\rm FS}(\nu)|\right)$. This Figure illustrates the effect of maser saturation at $\tau_0 \gsim 10$, at an amplification factor $\gsim 10^5$.

\begin{figure*} 
\vbox{\centerline{\epsfig{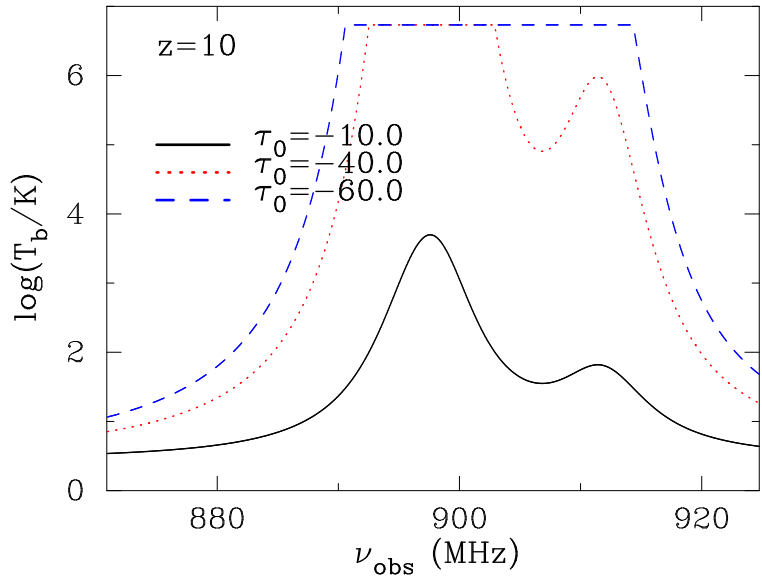}}}
\caption[]{The brightness temperature contrast, $\Delta T_{\rm b}(\nu)$, of the redshifted 3-cm transition of atomic hydrogen for three values of $\tau^{\rm FS}_0$ as a function of observed frequency $\nu_{\rm obs}$ in MHz. The inverted level populations of this transition give rise to stimulated emission that is orders of magnitude brighter than then CMB. The stimulated emission starts affecting the level populations for maser implication factors in excess $\sim 10^5$, and the maser saturates at $\Delta T_{\rm b}(\nu) \gsim 10^{6-7}$ K. The features in the spectrum are due to hyperfine splitting in the 3-cm transition. In the $\tau^{\rm FS}_0=-60$ case ({\it the blue dashed line}) maser saturation washes out this structure.}
\label{fig:signal}
\end{figure*} 
 {\it Hyperfine} splitting of the 3-cm line separates the $2p_{3/2}\rightarrow 2s_{1/2}$ transitions into three components at $\nu_1=9852$ MHz, $\nu_2=9876$ MHz, and $\nu_3=10030$ MHz, with line strength rations 1:5:2 \citep{Wild52,Ershov87,Dennison05}. Our previous calculations summed over all these transitions. However, these transitions are sufficiently separated in frequency that they can be resolved. We rewrite the expression for the brightness temperature contrast as

\begin{equation}
\Delta T_{\rm b}(\nu)=T_{\rm CMB}\left(\exp\left[\frac{|\tau^{\rm FS}_0|\phi(\nu)}{\phi(\nu_{\rm FS})}\right] - 1\right), 
\end{equation} where $\phi(\nu)$ is the 3-cm profile, and $\phi(\nu_{\rm FS})$ denotes the fine structure line evaluated at line centre. To account for hyperfine splitting, we approximate $\phi(\nu)$ as the sum over 3 Voigt profiles
\begin{equation}
\phi(\nu)=\frac{1}{8}\phi_1(\nu)+ \frac{5}{8}\phi_2(\nu)+\frac{2}{8}\phi_3(\nu),
\end{equation} where $\phi_n(\nu)$ denotes the Voigt profile of the $n^{\rm th}$ hyperfine transition. These profiles have the same spectral shapes, but are shifted in frequency.

Figure~\ref{fig:signal} shows $\Delta T_{\rm b}(\nu)$ as a function of $\nu$ for three different values of $\tau_0^{\rm FS}$. We have redshifted the profiles from $z=10$ into the observers frame. Note that the only way redshift entered our calculations is through the CMB-induced transitions. This Figure shows how stimulated emission amplifies the CMB enormously. The {\it black solid line} shows $\Delta T_{\rm b}(\nu)$ for $\tau_0=-10$ which boosts the CMB by a factor of $e^{10} \sim 10^4$, and clearly shows the hyperfine structure of the 3-cm transition. The {\it red dotted line} corresponds to the case of $\tau_0^{\rm FS}\sim -40$, which reaches a maximum at $\Delta T_{\rm b}(\nu) \sim 10^7$ K. For larger $\tau_0$ even the weaker hyperfine transition at $\nu_{\rm rest}=10030$ MHz is saturated, and the structure of the line is washed out (as illustrated by the {\it blue dashed line}).

The signal we computed is confined to a very small angular scale. The angular diameter of the cloud at $z=10$ is
\begin{equation}
\theta_{\rm cl}=\frac{2R_{\rm cl}}{D_A(z)}=1.2\left( \frac{n}{10^6\hs{\rm cm}^{-3}}\right)^{-1/3}\left( \frac{M}{10^7\hs M_{\odot}}\right)^{1/3}\hs {\rm mas}, 
\label{eq:mas}
\end{equation} where $d_{\rm A}(z)$ denotes the angular diameter distance to redshift $z$. Moreover, stimulated emission can be highly beamed \citep[e.g.][]{GK72,Alcock85a,Elitzur90a}, which would further reduce the angular extend of the high brightness temperature source. This beaming depends on the degree of saturation of the gas cloud: gas on the edge of a spherical cloud that amplifies radiation in a uniform and isotropic background (as is the case here), sees maximally amplified radiation coming from the opposite side of the cloud. Similarly, gas in the center of the cloud sees the lowest level of amplification. There are three degrees of saturation, which relate to the extend of the `unsaturated core' of the masing cloud \citep[see][]{GK72,Alcock85a,Alcock85b,Elitzur90a,Elitzur90b}.

\begin{itemize}[leftmargin=*]
\item In {\it unsaturated} clouds, maser amplification is not saturated in any part of the cloud. In our case, this means that the $\tau^{\rm FS}_0 \lsim 10$ (see Fig~\ref{fig:amp}). For these clouds, the CMB is exponentially amplified along all paths through the cloud.

\item In {\it partially saturated clouds}, the maser amplification is saturated outside a central region. In our case this translates to $10 \lsim \tau^{\rm FS}_0 \lsim 20$. The CMB is only amplified exponentially along paths that intersect the unsaturated core. The amplified CMB would point mostly radially outward outside the unsaturated core \citep[e.g.][]{GK72}, in which case the angular extend of the cloud is more closely related to the size of the unsaturated core.

\item In {\it full saturated clouds}, maser amplification is saturated everywhere. The angular size of the masing region is significantly smaller than the angular size of the cloud \citep{GK72,Elitzur90a,Elitzur90b}.
 
\end{itemize}

\begin{figure} 
\vspace{-5mm}
\vbox{\centerline{\epsfig{file=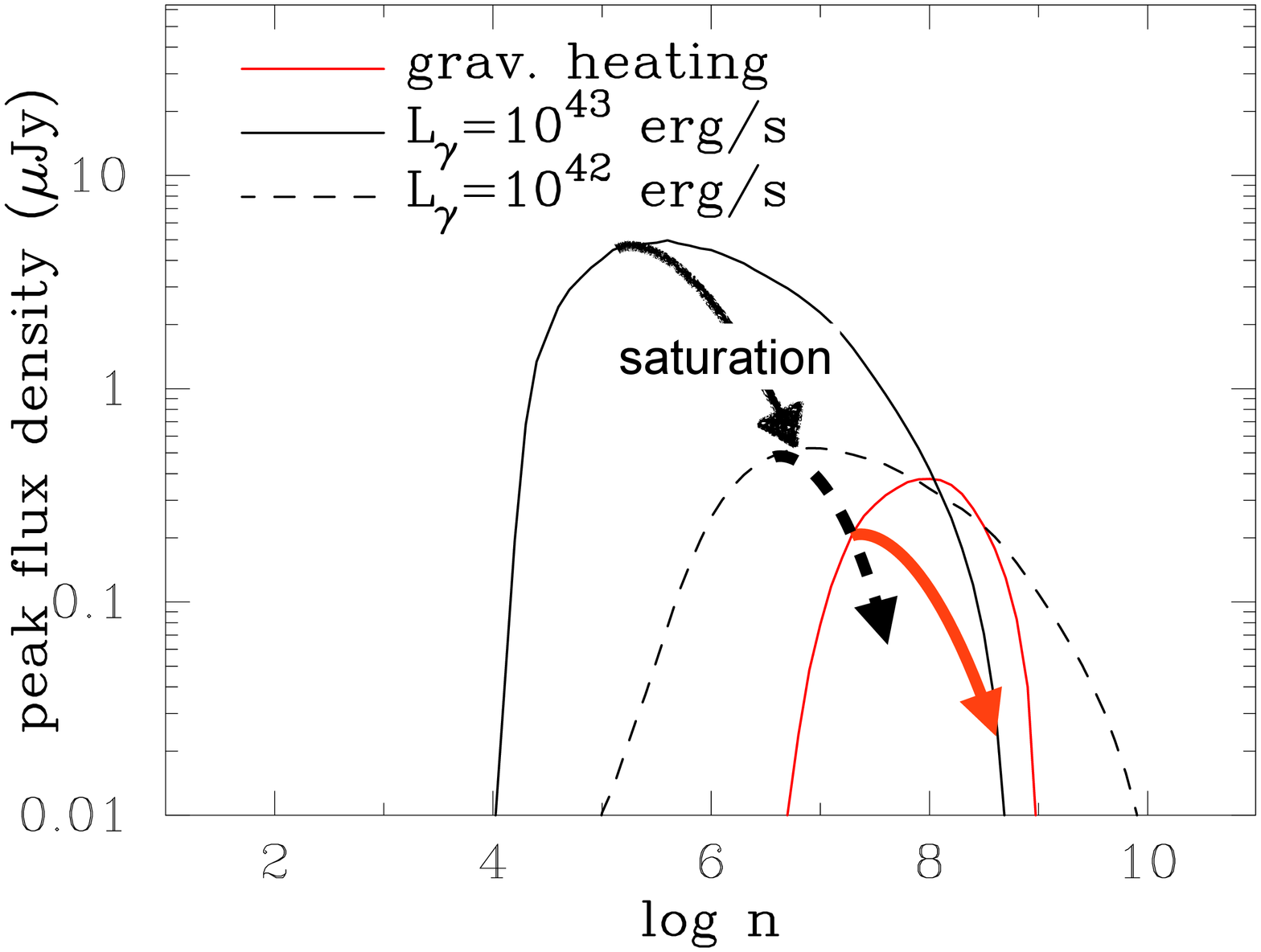,angle=270,width=10.5truecm}}}
\caption[]{The predicted flux density (in $\mu$Jy) in the $3$-cm transition as a function of cloud density $n$, for the three models discussed in previous plots. This Figure shows that for gravitational heating only, we reach a maximum flux density $F_\nu \sim 0.5$ $\mu$Jy for a narrow range of densities. This range is enhanced to $F_\nu \sim 5$ $\mu$Jy for the radiative heating model with $H^{\gamma}=10^{43}$ erg s$^{-1}$, and over a broader range of densities. These calculations assume that the masing cloud is unsaturated. This assumption breaks down at the densities that are indicated by the {\it thick arrows}, which indicate that the {\it thin lines} overpredict the flux densities. These predicted flux densities are within reach of surveys being planned with SKA1-MID (see text).}
\label{fig:fluxvsden}
\end{figure}

Angular sizes of spherical, partially and fully saturated masers have been calculated by \citet{GK72,Elitzur90a}. Results of these calculations cannot be applied directly to our results because in the 3-cm both the life-time and pumping rates of both levels participating in the maser differ greatly, while \citet{GK72} assumed equal life-times for both levels. We will defer modifying their formalism to future work, and will only focus on the unsaturated regime. We will indicate where our calculations break down. The observed flux density of the signal at a (observed) frequency $\lambda_0$ is
\begin{equation}
F_\nu = \frac{2 k_B}{\lambda_0^2 D^2_{A}(z)}\int_0^R 2\pi x \hs dx  \hs \Delta T_{\rm b}(\nu,x),
\label{eq:flux}
\end{equation} where $x$ denotes the projected distance from the center of the cloud. For $z = 10$ we have $\lambda_0 \simeq 33 \, \rm cm$. The {\it thin lines} in Figure~\ref{fig:fluxvsden} shows the predicted flux density (in $\mu$Jy) as a function of cloud density $n$ for each of the three models shown previously (as given by Eq~\ref{eq:flux}). Figure~\ref{fig:fluxvsden} shows that for the gravitational heating and radiative heating models with $H^{\gamma}=10^{42}$ erg s$^{-1}$ the model flux density peaks at $F_\nu \sim 0.5$ $\mu$Jy for a narrow range of densities. Radiative heating at $H^{\gamma}=10^{43}$ erg s$^{-1}$ can boost the flux up to $F_\nu \sim 5$ $\mu$Jy. The {\it thicker lines with arrows} indicate the density at which the cloud becomes partially saturated, and that Eq~\ref{eq:flux} overestimates $F_{\nu}$ at larger densities. 

Beaming is more pronounced in non-spherical masers. For flattened clouds, stimulated emission is highly beamed along longest axis of the cloud \citep[see][]{GK72,Alcock85a,Alcock85b}, and their apparent angular extend is more closely related to the physical area of the cloud perpendicular to this sightline. An additional complication to beaming through non-spherical clouds arises in the radiative heating models as these contain a central ionized bubble, which also affects the radiative transfer of stimulated emission. This effect is most important for the radiative heating model with $H^{\gamma}=10^{43}$ erg s$^{-1}$, where the radius of the ionized bubble at the predicted peak flux density is $\sim 10\%$ of that of the cloud as a whole (see Eq~\ref{eq:columnH}). This effect is likely much less important than the effects of beaming. Future calculations should nevertheless consider the impact of the ionized bubble.

The predicted 3-cm flux peaks at the onset of maser saturation. Here, we verify whether we were justified to ignore destruction of Ly$\alpha$ photons in the maser cycle in our calculation of the Ly$\alpha$ pumping rate. We focus on the radiative heating model with $H^{\gamma}=10^{43}$ erg s$^{-1}$, but note that the same reasoning applies to the other models. The predicted flux density peaks at $\log [n/{\rm cm}^{-3}] \sim 5$ for this model, at which $\tau^{\rm FS}_0\sim 10$. Stimulated emission from the amplified CMB occurs at a rate $\frac{\Omega_{\rm Amp}}{4\pi} \times {\rm Amplification} \times \Gamma^{\rm CMB}_{2p2s}\sim \frac{\Omega_{\rm Amp}}{4\pi}\times 0.3([1+z]/11)$ s$^{-1}$ for the associated amplification of $\sim 10^4$ (see Fig~\ref{fig:amp}). For the unsaturated case, most gas in the collapsing cloud sees a boost in the CMB well below this, as the maximum path length towards most gas is less than $\tau_0 \sim 10$. Only the gas on the edge of the cloud `sees' the maximum amplification, and only from a solid angle $\Omega_{\rm Amp} \ll 4\pi$. The destruction probability of a Ly$\alpha$ photon due to stimulated emission from the $2p$ state is the $P=\frac{\Omega_{\rm Amp}}{4\pi}\times 5\times 10^{-10}([1+z]/11)$, which limits the number of scattering events to $N_{\rm scat}\sim 2\times 10^{9}\frac{4\pi}{\Omega_{\rm Amp}}(11/[1+z])$, which is a factor $\sim 5 \frac{\Omega_{\rm Amp}}{4\pi}([1+z]/11)$ below the actual number of scattering events. Given that this correction only becomes relevant near the edge of the cloud, where $\Omega_{\rm Amp} \ll 4\pi$ (and where therefore this correction is small anyway), our overall predicted flux is barely affected by Ly$\alpha$ destruction in the maser cycle, and confirms that this effect becomes important at the onset of maser saturation.

Finally, we point out that free-free opacity $\kappa(\nu)=3.3\times 10^{-7}n^2_eT^{-1.35}_{e,4}(\nu/{\rm Ghz})^{-2.1}$ pc$^{-1}$ \citep{Condon92} is clearly negligible inside the collapsing cloud (here, $T_{e,4}$ denotes the electron temperature in units of $10^4$ K). In addition to this, the free-free opacity of the ionized IGM and our own Galaxy are negligible for $\nu \gsim 100$ Mhz (where $\nu$ denotes the frequency in the observer frame, see Spaans \& Norman 1997), which corresponds to $z \lsim 100$ for 3-cm masers. We discuss the detectability of this signal in \S~\ref{sec:detection} below.

\section{Detectability of the Signal}
\label{sec:detection}

The upcoming radio interferometer SKA will have the capability to detect the predicted
signal. As a part of the Key Science Projects (KSP) with SKA1-MID (the first
phase of SKA at medium frequencies: $0.4 \, {\rm GHz} < \nu < 20 \, \rm GHz$) thousands
of hours of integration will be carried out in the frequency range of interest to  us for line (redshifted HI) and continuum surveys, with ultra
deep surveys using integration of over 1000~hours on a  single pointing\footnote{see e.g. http://astronomers.skatelescope.org/wp-content/uploads/2015/08/SKA-SCI-LVL-001W.pdf}. Such  ultra-deep surveys might carry out up  to 2~years of integration on multiple pointings. Planned  continuum surveys with bandwidth $\Delta\nu = 0.3 \nu$ will reach RMS noise of $100 \, \rm nJy$
in  1000~hours at $\nu \simeq 900 \, \rm MHz$. Figure~\ref{fig:signal} shows the line width of 
the  2p-2s signal $\delta\nu \simeq 10\hbox{--}25 \, \rm MHz$. Assuming 
$\delta\nu = 10 \, \rm MHz$  gives an RMS of nearly $500 \, \rm nJy$ at $\nu \simeq 900\, \rm MHz$, which 
will allow for a significant detection for some of our models using SKA1-MID. 

What is the expected number of observable 2p-2s masers from $z\simeq 10$? The space density of DCBHs at $z \simeq 10$ is highly uncertain, and could lie in the range $10^{-3} \hbox{--}10^{-10} \, \rm cMpc^{-3}$ \citep[see e.g. Fig~4 of][]{D14}. The comoving volume corresponding to a single SKA pointing (angular extent of nearly 1~square degree and bandwidth of 300~MHz for a central frequency $\simeq 900 \,\rm MHz$) is $\simeq 2\times 10^8 \, \rm cMpc^{3}$, covering a redshift range $8 <z < 12$. Even though a single pointing might contain a large number of DCBHs, observing their 3-cm fine structure transition is challenging, as they are observable only when they are emitting at peak flux, which occurs only for a limited range of densities.  

\section{Discussion of Model Assumptions}
\label{sec:discuss}

\begin{enumerate}[leftmargin=*]

\item The number density $n_{\rm 2p}$ depends directly on $\langle N_{\rm scat} \rangle$. We adopted $\langle N_{\rm scat} \rangle=k_1 \tau_0$ which is strictly only valid for a static medium. The clouds we are interested in are collapsing, and we may therefore have overestimated $\langle N_{\rm scat} \rangle$. The effect of ignoring the kinematics is not important. As we mentioned earlier, the collapse of the clouds occurs at velocities of order the circular velocity of the dark matter halo, i.e. $v\sim 10$ km s$^{-1}$, which corresponds to the thermal width of the line profile. The velocity gradients through the clouds are therefore of order $v/R_{\rm cl}$, and the Sobolev optical depth is $\tau \sim v_{\rm th}/[v/R_{\rm cl}]\sim R_{\rm cl}$. In other words, velocity gradients only introduce a minor correction to the scattering rate\footnote{\citet{Bonilha79} show that $\langle N_{\rm scat} \rangle$ reduces to $\langle N^{\rm vel}_{\rm scat} \rangle$ by an amount $\frac{\langle N^{\rm vel}_{\rm scat} \rangle}{\langle N_{\rm scat} \rangle}=\frac{1}{1+0.05 \xi^{1.5}}$, where $\xi =\frac{1}{\sqrt{3}}B\frac{\Delta V}{x_{\rm peak}}$, in which $B=(a_v\tau_0/\sqrt{\pi})^{1/3}$ is the boost in path length that we introduced before (see \S~\ref{sec:level}). $\Delta V$ denotes the difference in velocity between cloud edge and centre, and $x_{\rm peak}$ is the frequency at which Ly$\alpha$ photons are most likely to escape. For the column densities of interest, $x_{\rm peak}> 10^3$ km s$^{-1}$ (see Eq~ in Dijkstra 2014), $B \sim 10^2$, and we get that $\xi$ is of order unity. It is then clear that velocity gradients do not affect $\langle N_{\rm scat} \rangle$ at all in our scenario.}. It is also worth stressing that $\langle N_{\rm scat} \rangle$ is limited in our model by collisional deexcitation and photoionisation from the $n=2$ level, and as a result Ly$\alpha$ photons typically diffuse in real space by a fraction of the physical size of the cloud (also see discussion below).

Of course, given that $\Delta T_{\rm b}(\nu)$ depends exponentially on $\langle N_{\rm scat} \rangle$ it is worth studying this effect in more detail for more realistic models. One other aspect that can be addressed with more realistic models is that even though fragmentation is suppressed, density inhomogeneities can possibly give rise to lower column densities paths which allow Ly$\alpha$ photons to escape, and which can suppress $\langle N_{\rm scat} \rangle$. These calculations are challenging as ordinary techniques that are used to perform Ly$\alpha$ radiative transfer simulations in simulations often speed-up the transfer problem by skipping the majority of scattering events.

\item Our adopted $\langle N_{\rm scat} \rangle$ is appropriate for a static, spherical gas cloud. For flattened gas clouds, it is likely that Ly$\alpha$ photons escape in the direction of lowest $N_{\rm HI}$, which can reduce $\langle N_{\rm scat} \rangle$ at fixed $n$. However, Figure~\ref{fig:nscat} in the Appendix shows clearly how for both the gravitational and radiative heating cases, $\langle N_{\rm scat} \rangle$ is suppressed by $\sim 2-3$ orders of magnitude due to collisional deexcitation and photoionisation from the $n=2$ level compared to calculations that ignore these processes. This suggests that each Ly$\alpha$ photon is effectively destroyed after it traversed a column density that is $\sim 2-3$ orders of magnitude smaller than the cloud column density (since $\langle N_{\rm scat} \rangle \propto N_{\rm HI}$). In other words, Ly$\alpha$ photons have only moved a fraction of the physical size of the cloud before being destroyed. This implies that $\langle N_{\rm scat} \rangle $ depends weakly on the assumed spherical geometry.

\item We also ignore velocity structure in the cloud when we compute the maser amplification factor. This assumption is again safe. The  profile of the fine structure lines are extremely broad \citep{Wild52,Ershov87,Dennison05}. For example, the absorption line profile falls by a factor of ~2 at nearly $\sim 10^3$ km s$^{-1}$ away from line centre (and $\phi(\nu)$ drops only by $1\%$ 100 km s$^{-1}$ away from line center). Gas motions can therefore be safely ignored.

\item Throughout our analysis we always assumed that the $2p_{3/2}$ and $2p_{1/2}$ levels were populated according to their statistical weight, i.e. $n_{{\rm 2p}_{3/2}}=2n_{\rm 2p}/3$. Both levels are populated by Ly$\alpha$ scattering. The frequency off-set between $2p_{3/2}$ and $2p_{1/2}$ is $\Delta \nu \sim 10^{10}$ Hz, which is $\sim 10\%$ of the thermal width of the Ly$\alpha$ line at T$\sim 10^4$ K (i.e. $\Delta \nu \sim 0.1 \Delta \nu_{\rm D}$). We expect the rate at which Ly$\alpha$ photons to populate the $2p_{3/2}$ and $2p_{1/2}$ levels to depend on the shape of the Ly$\alpha$ spectrum around these frequencies. In other words, we expect the rate at which the $2p_{3/2}$ and $2p_{1/2}$ are populated to depend on the color-temperature of the Ly$\alpha$ radiation field near the line resonance, which equals the gas temperature when gas is extremely opaque to Ly$\alpha$ radiation \citep{Wouthuysen52,Field58}. Because the mean thermal energy of the gas ($kT$) exceeds the energy difference between the $2p_{3/2}$ and $2p_{1/2}$ levels, we expect these two levels to be in statistical equilibrium.

\end{enumerate}

\section{Conclusions}
\label{sec:conc}
The direct collapse black hole (DCBH) scenario describes the isothermal collapse of a pristine gas cloud directly into a massive, $M_{\rm BH}=10^4\hbox{--}10^6 M_{\odot}$ black hole. DCBH formation is a remarkably complex problem to tackle theoretically, and it would be extremely helpful to have observational diagnostics of this process.

In this paper, we have studied the detectability of the {\it fine structure} transitions of atomic hydrogen from gas clouds collapsing into or onto a DCBH. We have focussed on the strongest fine-structure transitions, namely the $2s_{1/2}\hbox{--}2p_{3/2}$ transition with a rest frame wavelength of $\lambda=3.04$ cm, and the $2p_{1/2}\hbox{--}2s_{1/2}$ transition at $\lambda =27 $ cm. A detectable fine-structure signal from atomic hydrogen thus requires a non-negligible population of hydrogen atoms to be in the first excited ($n=2$) state. It has long been realised that Ly$\alpha$ scattering in optically thick gas can enhance especially the $2p$-level of HI \citep[e.g.][]{Pottasch}, which can lead to inverted level populations in the $2s_{1/2}-2p_{3/2}$ transition \citep[][]{FP61}. Observational searches for 3-cm maser activity from nearby HII regions have not been successful, because Ly$\alpha$ pumping of hydrogen in nearby HII regions is not effective enough to give rise to sufficient atomic hydrogen in its excited state.

 In this paper we have shown that large HI column densities of primordial gas at $T\sim 10^4$ K, combined with a low molecular hydrogen abundance---which represent key requirements in the DCBH scenario---provide optimal conditions for pumping of the $2p$-level of atomic hydrogen by trapped Ly$\alpha$ photons. We show that simplified models of the DCBH scenario give rise to a minimum optical depth through the 3-cm line, $\tau_{\rm 3cm} \sim-40$. We show that these models predict that CMB radiation passing through a cloud directly collapsing into a DCBH is amplified by up to a factor of $\sim$ $10^5$. For larger amplification factors the amplified CMB affects the $2p$ and $2s$ level populations such that further amplification is halted, and the maser saturates. 
 
 Hyperfine splitting of the 3-cm transition gives rise to a characteristic broad (FWHM$\sim$ tens of MHz in the observer's frame), asymmetric line profile, which is insensitive to the gas kinematics. The predicted signal subtends a small angular scale of $\sim 1-10$ mas, which translates to a the peak flux density in the range $0.3\hbox{--}3 \, \rm \mu Jy$ with a line width of $\delta\nu \simeq 20 \, \rm MHz$. This signal can be detected by using the spectral cube data from already-planned ultra-deep continuum and line surveys with SKA1-MID. This is remarkable, as it implies it may be possible to directly detect gas in emission in high-redshift ($z\sim 10-20$) from {\it individual} atomic cooling dark mater halos.
 
CR7, a recently discovered unusually luminous Ly$\alpha$ emitting source at $z\sim 6.6$ \citep{Sobral15}, has been argued to be the first DCBH candidate \citep[see][]{Sobral15,PF15,Ag15}. The large Ly$\alpha$ luminosity of CR7 implies a high escape fraction of Ly$\alpha$ photons. In contrast, we have shown that Ly$\alpha$ photons are destroyed through collisional processes when the 3-cm maser signal is maximized. Furthermore, the observed width of the Ly$\alpha$ spectral line (FWHM$\sim 266$ km s$^{-1}$, see Sobral et al. 2015) indicates that resonant scattering of Ly$\alpha$ photons - which broadens the Ly$\alpha$ spectral line (see Dijkstra 2014, and references therein) - is inconsistent with a scenario in which Ly$\alpha$ scatter $\log \langle N_{\rm scat} \rangle \gsim 8$ (see Dijkstra \& Gronke, in prep). This implies that if CR7 is indeed associated with a DCBH, then it must represent an evolutionary stage during which there is no (detectable) stimulated 3-cm emission. 
 
 Our results were clearly obtained from a simplified representation of the DCBH scenario, which allowed us to focus entirely on identifying and modelling the relevant radiative processes. We stress that it is important to study these radiative processes in more realistic gas distributions especially because gas geometry can strongly affect the beaming of the maser, which affects the apparent angular scale of the masing cloud and therefore the predicted maser flux.
 
 While challenges remain on the modelling side, we stress that it is well worth addressing these in future work, as the masing conditions that we found are uniquely associated with the physical conditions that enable the DCBH scenario: these conditions include chemically pristine gas (i.e. no dust) inside dark matter halos with $T_{\rm vir} \sim 10^4$ K, in which molecular hydrogen formation and gas fragmentation have been suppressed. These last two additional requirements are important: ordinary pristine gas inside an atomically cooling halo would form H$_2$ during its collapse in quantities that are fatal for the required pumping of the maser levels. Similarly, once gas is allowed to fragment and clump, the radiative transfer of Ly$\alpha$ proceeds differently, with Ly$\alpha$ photons preferentially escaping through lower column density holes and scattering less frequently \citep[see e.g.][]{Neufeld91,Haiman99,GD14}. The Ly$\alpha$ pumping efficiency is reduced in pristine environments {\it not} associated with the DCBH scenario. This implies that a detection of the redshifted 3-cm signature in deep SKA surveys would provide direct and unique evidence for the formation of supermassive black holes via the direct collapse of a gas cloud.
  
{\bf Acknowledgements} MD thanks for the Raman Research Institute for their hospitality during a visit which started this project. We thanks Jonathan Pritchard, Max Gronke, Lluis Mas-Ribas for useful discussions, and Jens Chluba for helpful correspondence. This work was supported in part by NSF-grant AST-1312034 (for AL). We thank an anonymous referee for an excellent, constructive report.

\label{lastpage}

\appendix
\section{Verifying the Dominance of Stimulated Emission}
\label{app}

We approximated Eq~\ref{eq:RTsolution}
\begin{equation}
I_{\nu}(s)=I_{\nu,0}{\rm e}^{-\kappa_{\nu} s}+\frac{j_{\nu}}{\kappa_{\nu}}\left(1- {\rm e}^{-\kappa_{\nu} s}\right)\approx I_{\nu,0}{\rm e}^{-\kappa_{\nu} s},
\end{equation} which is valid only when
\begin{equation}
I_{\nu,0}\gg \left|\frac{j_{\nu}}{\kappa_{\nu}}\right|.
\label{eq:comp1}
\end{equation} This section shows this condition is generally met throughout our calculations. If we substitute the definitions of $\kappa_{\nu}$ and $j_{\nu}$ (see Eq~\ref{eq:kj}) we get
\begin{equation}
\frac{j_{\nu}}{\kappa_{\nu}}=\frac{2\nu^3 h}{c^3\left( \frac{g_u}{g_l}\frac{n_l}{n_u}-1\right)}=\frac{2E_{\rm FS}}{\lambda^2_{\rm FS}\left( \frac{g_u}{g_l}\frac{n_l}{n_u}-1\right)},
\label{eq:jk}
\end{equation} where in the second step we choose to evaluate this at line centre of the fine-structure transition (i.e. $\nu=\nu_{\rm FS}$, $\lambda=\lambda_{\rm FS}$). Eq~\ref{eq:comp1} can therefore be rephrased as 
\begin{equation}
\frac{2kT_{\rm CMB}}{\lambda^2_{\rm FS}}\gg \left|\frac{2 E_{\rm FS}}{\lambda^2_{\rm FS}\left( \frac{g_u}{g_l}\frac{n_l}{n_u}-1\right)}\right|,
\end{equation} which can be simplified to
\begin{equation}
\frac{kT_{\rm CMB}}{E_{\rm FS}}\gg \left|\frac{1}{\left( \frac{g_u}{g_l}\frac{n_l}{n_u}-1\right)}\right|
\label{eq:condition}
\end{equation} Since $T_{\rm CMB}=2.725(1+z)$ K, and $E_{\rm FS}/k \sim 0.5$ K for the 3-cm transition, and $\sim 0.05$ K for the 27-transition.When we have stimulated emission we have $n_u \gg n_l$, and the R.H.S is $1$. In case there is stimulated emission, the left-hand side is clearly much greater than the right hand side and our approximation for the solution of the radiative transfer equation is valid.

\section{Intermediate Quantities}
\label{app:intermediate}

\begin{figure}
\vbox{\centerline{\epsfig{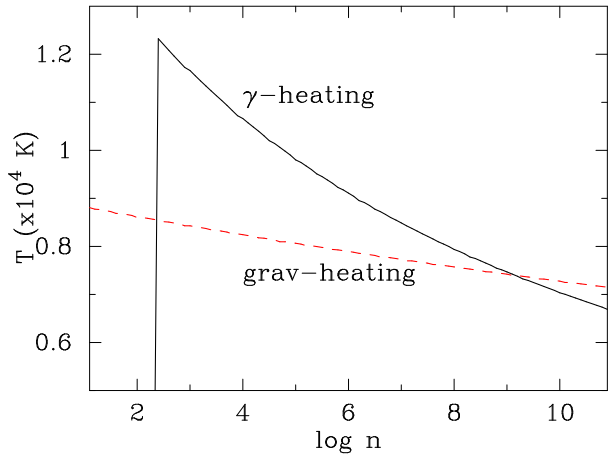}}}
\vspace{0mm}
\caption[]{The temperature evolution in the gravitational heating model ({\it red dashed line}), and radiative heating with $H^{\gamma}=10^{43}$ erg s$^{-1}$ ({\it black solid line}). In the radiative heating case, the total heating - and therefore cooling - rate is constant. As density increases the temperature decreases to maintain a constant total cooling rate. The density-dependence of $T$ in the gravitational heating model is weaker because the heating rate increases with density.} 
\label{fig:temp}
\end{figure}

\begin{figure}
\vbox{\centerline{\epsfig{file=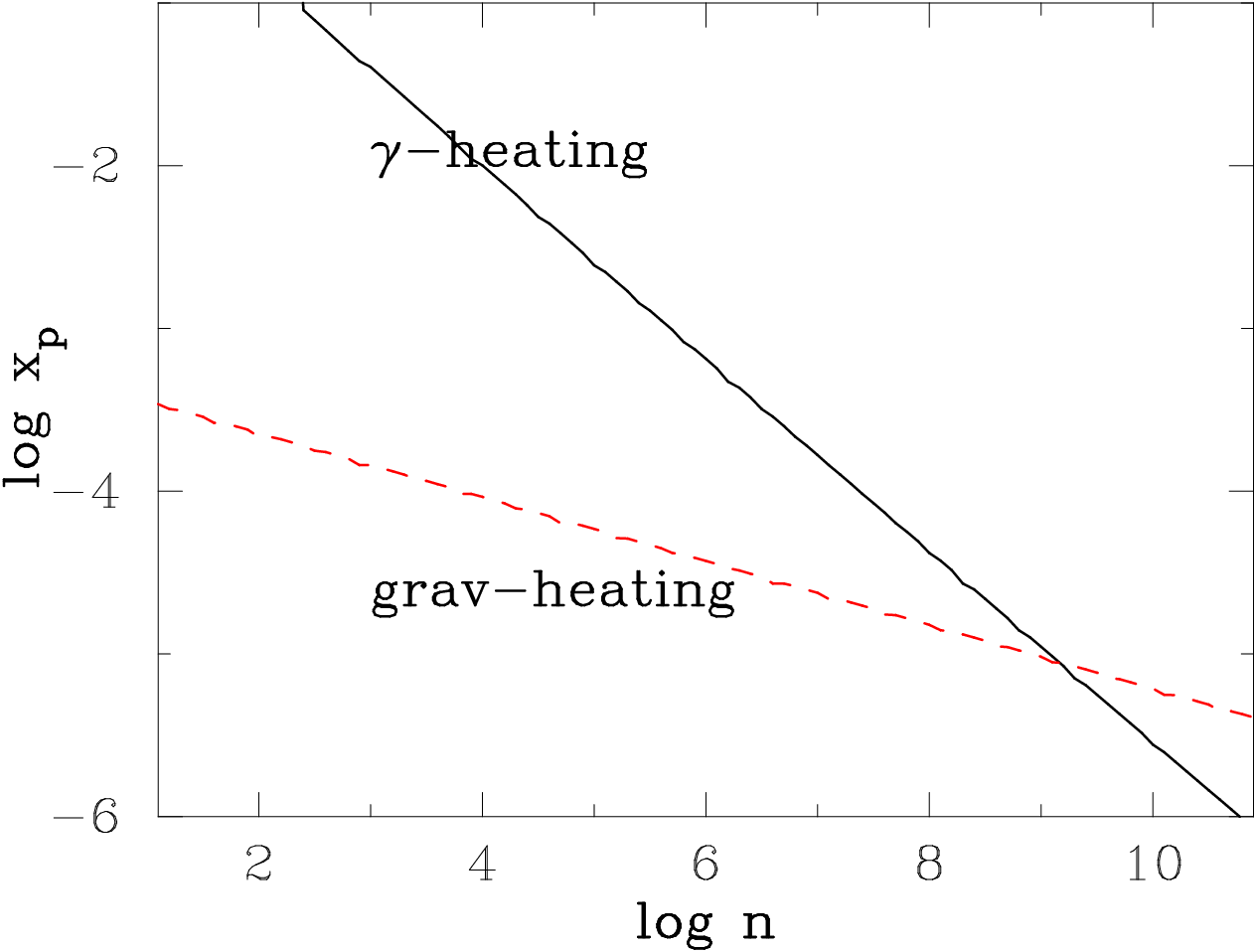,angle=270,width=8.0truecm}}}
\vspace{0mm}
\caption[]{Same as Fig~\ref{fig:temp}, but now we show the ionised fraction $x_{\rm p}=x_{\rm e}$. In the radiative heating case, the gas becomes fully ionised at $n < n_{\rm crit} \sim 240$ cm$^{-3}$ (see Eq~\ref{eq:ncrit}), and the ionised fraction goes to $1$.} 
\label{fig:xp}
\end{figure}

\begin{figure}
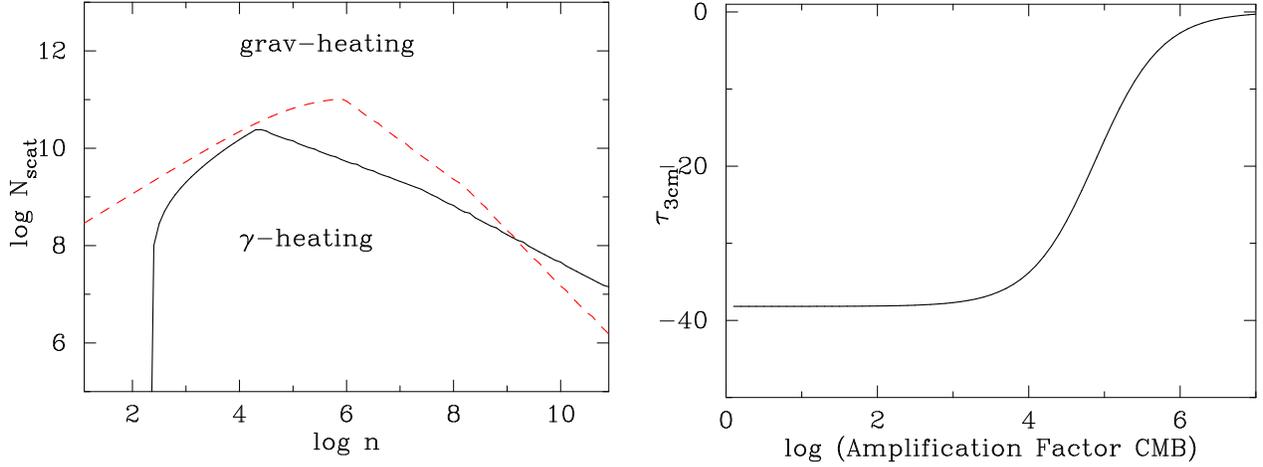

\vbox{\centerline{\epsfig{file=nscat.eps,angle=270,width=8.0truecm}\hspace{6mm}\epsfig{file=saturation.eps,angle=270,width=8.0truecm}}}
\vspace{0mm}
\caption[]{{\it Left:} Same as Fig~\ref{fig:temp}, but now we show $\langle N_{\rm scat} \rangle$. For both cases $\log \langle N_{\rm scat} \rangle\sim 8-11$. The quantitative behaviour of both curves is discussed in the text. We have also indicated at what the number density the fine structure optical depth reaches its minimum. {\it Right:} This plot shows the optical depth in the 3-cm transition as a function of CMB-amplification factor. Stimulated emission from the amplified CMB reduces the $2p$-level population, and hence the absolute value of $\tau_{\rm 3cm}$, only when the CMB is amplified more than a factor of $\sim 10^5$. Maser saturation is not important up until these amplification factors.} 
\label{fig:nscat}
\end{figure}

Figures~\ref{fig:temp}-\ref{fig:nscat} show intermediate results of our calculation. Figure~\ref{fig:temp} shows the density-dependence of $T$ for the gravitational heating model ({\it red dashed line}) and radiative heating model with $H^{\gamma}=10^{43}$ erg s$^{-1}$ ({\it black solid line}). In the radiative heating case, the cooling rate is constant, $L_{\rm cool}\propto n$ (see \S~\ref{sec:num2s2p}). To maintain a constant total cooling rate the temperature must decrease. The density-dependence of $T$ in the gravitational heating model is weaker because the cooling rate increases with density as $L_{\rm cool}\propto n^{5/3}$ (see \S~\ref{sec:num2s2p}).

Figure~\ref{fig:xp} shows the density-dependence of the associated ionised fraction $x_{\rm e}=x_{\rm p} \equiv n_{\rm p}/n$. The different density dependence of $x_{\rm p}$ between both models is driven entirely by the different temperature evolution. Because in the model with radiative heating the cloud is fully ionised for $n < n_{\rm crit} \sim 240$ cm$^{-3}$, the ionised fraction goes to unity.

Finally, the {\it left panel} of Figure~\ref{fig:nscat} shows the density-dependence of $\langle N_{\rm scat} \rangle$. For the gravitational heating model we have\footnote{The actual density dependence is slightly steeper than this, because $\tau_0 \propto N_{\rm HI}T^{-1/2}$, and $T$ decreases with $n$, albeit slowly.} $\langle N_{\rm scat} \rangle \propto \tau_0 \propto n^{2/3}$. For $\log n \gsim 5.6$, collisional deexcitation limits $\langle N_{\rm scat} \rangle$ and it decreases as $\langle N_{\rm scat} \rangle \propto n_{\rm p} ^{-1}\propto n^{-0.8}$. For the model with radiative heating $\langle N_{\rm scat} \rangle=0$ for $n < n_{\rm crit}$ as the gas is fully ionised. It rises to catch up with $\langle N_{\rm scat} \rangle$ for the gravitational model, as the column densities become increasingly similar for both models as $n$ increases above $n> n_{\rm crit}$. However, because of the larger ionised fraction in the radiative heating models, collisional deexcitation becomes important at lower density and $\langle N_{\rm scat} \rangle$ decreases. $\langle N_{\rm scat} \rangle$ is equal for both models at the density where the gas has the same $T$.

\section{Maser Saturation}
\label{app:saturation}

As the CMB is amplified by a factor of $B_{\rm CMB}$ through the cloud collapsing into/onto a DCBH, the CMB induced transitions between the $2p$ and $2s$ levels increase by this same factor. Our original calculations did not account for this boost. Here, we repeat our calculations for the radiative heating model with $H^{\gamma}=10^{43}$ erg s$^{-1}$, at a fixed number density $\log n \sim 7$, but boost the CMB-induced radiative transitions by a factor of $B_{\rm CMB}$, i.e. $\Gamma^{\rm CMB}_{2s2p} \rightarrow B_{\rm CMB}\Gamma^{\rm CMB}_{2s2p}$ and $\Gamma^{\rm CMB}_{2p2s} \rightarrow B_{\rm CMB}\Gamma^{\rm CMB}_{2p2s}$. The {\it right panel} of Figure~\ref{fig:nscat} shows $\tau_{{\rm 3cm},0}$ as a function of $B_{\rm CMB}$. This plot shows that stimulated emission by the CMB starts affecting the $2p$-levels only when $B_{\rm CMB} \sim 10^5$, because only then does the de-population rate via stimulated emission become comparable to the population rate through Ly$\alpha$ scattering. 

\section{Some Useful Numbers}
\label{app:numbers}
We can verify the magnitude of our calculations by comparing the cross-section for stimulated $3-$cm transition to that of some known electronic transition. We generally have
\begin{equation}
\kappa_{\nu}=\kappa(\nu)=\frac{h\nu_{\rm ul}B_{\rm ul}}{4\pi  \Delta \nu_{\rm ul}}\times \Big{(}\frac{g_u}{g_l}n_{\rm l}-n_{\rm u} \Big{)}\phi(\nu).
\end{equation} We will now evaluate the cross-section in some transitions based on this.
\begin{enumerate}
\item For Ly$\alpha$ we have $n_{u} \ll n_l$, $g_u=3$, $g_l=1$, $A_{ul}=6.25 \times 10^8$ s$^{-1}$, and $\phi(\nu_{\alpha})=1/\sqrt{\pi}$. This gives us 
\begin{equation}
\sigma_0=\frac{\kappa(\nu_{\alpha})}{n_{1s}}=\frac{3}{8\pi}\frac{\lambda^2_{\alpha}A_{\alpha}}{\sqrt{\pi}\Delta \nu_{\alpha}}=5.89\times 10^{-14}(T/10^4)^{-1/2}\hs{\rm cm}^{-2},
\end{equation} which is a familiar result (see e.g. Dijkstra 2014).

\item For H$\alpha$ the velocity averaged cross-section can be computed similarly. However, we take a short cut and point out that
\begin{equation}
\sigma_{0,{\rm H}\alpha}=\sigma_0\left(\frac{f_{{\rm H}\alpha}\lambda_{{\rm H}\alpha}}{f_{{\rm Ly}\alpha}\lambda_{{\rm Ly}\alpha}}\right)=4.86\times 10^{-13}(T/10^4)^{-1/2}\hs{\rm cm}^{-2},
\end{equation} where $\lambda_{{\rm Ly}\alpha}=1215.67$ \AA, $\lambda_{{\rm H}\alpha}=6562.8$ \AA, and the oscillator strengths $f_{{\rm Ly}\alpha}=0.416$ and $f_{{\rm H}\alpha}=0.637$.

\item For stimulated 3-cm Eq~\ref{eq:tau} shows that (for $n_u \gg n_l$)
\begin{equation}
\sigma_{3{\rm cm},0}=-\frac{\lambda^2A_{\rm ul}}{2\pi A_{\alpha}}=-2.0 \times 10^{-15}\hs {\rm cm}^{2},
\end{equation} 
\end{enumerate} where the negative sign reflects the negative opacity of the inverted level population. Comparing the last two numbers we get 
\begin{equation}
\frac{\tau_{0,{\rm H}\alpha}}{\tau_{3{\rm cm},0}}=\frac{3}{2}\frac{\sigma_{0,{\rm H}\alpha}}{\sigma_{3{\rm cm},0}}\approx -362(T/10^4\hs{\rm K})^{-1/2}
\end{equation} where the factor $3/2$ represents the fraction of atoms in the $2p$ state that is in the $2p_{3/2}$ state. We note that this 
is a factor of $\sim \sqrt{2}$ lower than the number given in \citet{FP61}, who find that $\frac{\tau_{0,{\rm H}\alpha}}{\tau_{3{\rm cm},0}}\sim -700$ for $T=5000$ K (they adopt $v_{\rm th}=9.1$ km s$^{-1}$). The origin of this (small) difference is unclear.

\section{Ly$\alpha$ Destruction by Molecular Hydrogen}
\label{app:h2}
\citet{Neufeld90} provides an expression for the escape fraction of Ly$\alpha$ photons from a slab of gas with center-to-edge column density $N_{\rm HI}$, and that contains a molecular hydrogen fraction $f_{H_2}\equiv n_{H_2}/n_{\rm H}$. This expression is
\begin{eqnarray}
f_{\rm esc}^{H2}(\tau_0)=\frac{4}{\pi} \sum_{n=1}^{\infty}\frac{(-1)^{n-1}[(2n-1)+(\sqrt{6}\epsilon_a\tau_0/\pi)(1-x_n)]}{A_n} \nonumber, \\
A_n=(2n-1)[(2n-1)+\sqrt{6}(\epsilon_a+\epsilon_0)\tau_0/\pi]+ (6\epsilon_a\epsilon_0\tau_0^2/\pi^2)(1-x_n^2)
\end{eqnarray}, where $x_n=\exp\left(-\pi (n-\frac{1}{2})\sigma_a/\tau_0\right)$, and
\begin{equation}
\sigma_a=8.3\times 10^5 T^{-1}_4,\hspace{5mm} \epsilon_0=0.082\times f(2,6)\times f_{H_2}, \hspace{5mm}\epsilon_a=0.069\times f(2,5)\times f_{H_2},
\end{equation} in which $f(2,5)$ and $f(2,6)$ denote the fraction of $H_2$ atoms in the $v=2, J=5$ state and $v=2, J=6$ state. We adopt $\log f(2,5)=-1.8$ and $\log f(2,6)=-2.4$ from Figure~19 of Neufeld (1990). These numbers assume LTE.

\section{Other Ly$\alpha$ Destruction Mechanism}
\label{app:other}

\begin{enumerate}[leftmargin=*]
\item Photoionization from the $2p$-level by the quasar: The radiative heating models contain a luminous source in the centre of the cloud. Hydrogen atoms in the $2p$ can be photoionized by photons with $E> 3.4$ eV, which can penetrate into the neutral gas. This photoionisation rate can be estimated from
\begin{equation}
\Gamma^{\rm 2p}_{\rm ion}=\dot{N}_{\rm 2p-ion}\frac{\langle\sigma_{\rm ion}\rangle }{4\pi r^2},
\end{equation} where $\dot{N}_{\rm 2p-ion}$ denotes the rate at which photons in the energy range $E=3.4-13.6$ eV are produced, $\langle\sigma_{\rm ion}\rangle$ denotes the frequency averaged cross-section. Substituting some numbers gives:
\begin{equation}
\Gamma^{\rm 2p}_{\rm ion}=0.1\left(\frac{\dot{N}_{\rm 2p-ion}}{10^{54}\hs {\rm s}^{-1}}\right)\left(\frac{r}{1 \hs {\rm pc}}\right)^{-2} \hs {\rm s}^{-1}
\end{equation} where we assumed that $\langle\sigma_{\rm ion}\rangle=1.4\times 10^{-17}$ cm$^{-2}$. For comparison, the cloud radius at the minimum $\tau^{0}_{\rm FS}$ is $\sim 3$ pc is $R_{\rm cl}\sim 2$ pc, at which $\log x_{\rm p} \sim -3.8$, and the collisional deexcitation rate is $C_{2p2s}n_{\rm p} \sim 0.5$ s$^{-1}$. That is, photoionization from the $2p$ state can be important in the inner parts of the cloud in the radiative heating case, but is not a show-stopper.

\item Photodetachment of $H^-$. Ly$\alpha$ photons can detach the electron from the H$^-$ ion. The cross-section for this process is $\sigma=5.9\times 10^{-18}$ cm$^{-2}$ \citep[e.g.][]{Sp87} for Ly$\alpha$ photons, which is almost an order of magnitude larger than the photoionisation cross-section from the $n=2$ level. So, unless the H$^{-}$ number density exceeds $0.1[n_{\rm 2p}+n_{\rm 2s}]$, we do not consider this process important. Some numbers, for radiative heating we reach the minimum in $\tau$ for $\log n \sim 7$, where we have $\log n_{\rm 2p}\sim -2.6$. We thus have a fractional number density of $\log n_{\rm 2p}/n\sim -9.6$, while one-zone models indicate a $H^-$ fraction a bit below $\log n_{H^-}/n \sim -11$ \citep[][]{Latif14}. In other words, photo detachment of $H^-$ is not negligible, but it is slightly less important that photoioization from the $n=2$ level, which is included in our calculations. Similar, for the gravitational heating model we have $\log n_{\rm 2p}/n \sim -10.7$ at the minimum for $\log n \sim 9$, while $\log n_{H^-}/n \sim -12$  \citep[][]{Latif14}. In addition, the destruction of Ly$\alpha$ photons via photoionization from the $n=2$ level is negligible compared to the destruction due to collisional mixing of the $2p$ and $2s$ levels at the densities where we have the strongest maser activity. 
\end{enumerate}

\end{document}